\documentclass[twocolumn,tighten]{aastex63}

\usepackage{hyperref}

\accepted{30 October 2020}

\shorttitle{The fast build-up of massive ETGs}
\shortauthors{Saracco et al.}

\begin{document}

\title{The Rapid Build-up of Massive Early-type Galaxies.\\ 
Supersolar Metallicity, High Velocity Dispersion and Young Age for an ETG at z=3.35.}

\correspondingauthor{Paolo Saracco}
\email{paolo.saracco@inaf.it}

\author[0000-0003-3959-2595]{Paolo Saracco}
\affil{INAF - Osservatorio Astronomico di Brera, via Brera 28, 20121 Milano, Italy}

\author{Danilo Marchesini}
\affiliation{Tufts University, Physics and Astronomy Department, 574 Boston Ave, Medford, 02155 MA, USA}

\author{Francesco La Barbera}
\affiliation{INAF - Osservatorio Astronomico di Capodimonte, sal. Moiariello 16, 
80131 Napoli, Italy}

\author{Adriana Gargiulo}
\affiliation{INAF - Istituto di Astrofisica e Fisica Cosmica, IASF, via A. Corti
 12, 20133 Milano, Italy}

\author{Marianna Annunziatella}
\affiliation{Centro de Astrobiologia (CSIC-INTA), Spain (PI) }

\author{ Ben Forrest}
\affiliation{Department of Physics and Astronomy, University of California, Riverside, CA 92521, USA}

\author{Daniel J. Lange Vagle }
\affiliation{Tufts University, Physics and Astronomy Department, 574 Boston Ave, Medford, 
02155 MA, USA }

\author{Z. Cemile Marsan}
\affiliation{Department of Physics and Astronomy, York University, Toronto, Ontario, Canada }

\author{Adam Muzzin }
\affiliation{Department of Physics and Astronomy, York University, Toronto, Ontario, Canada }

\author{Mauro Stefanon }
\affiliation{ Leiden Observatory, Leiden University, Leiden, Netherlands}

\author{ Gillian Wilson}
\affiliation{Department of Physics and Astronomy, University of California, Riverside, CA 92521, USA }



\begin{abstract}
How massive early-type galaxies assembled their mass, on which timescales 
the star formation quenched, when their supersolar metallicity has been established,
are still open and debated issues.
Thanks to very deep spectroscopic observations carried out at the Large Binocular Telescope, 
we measured simultaneously stellar age, metallicity and velocity dispersion for C1-23152,
an ETG at redshift $z$=3.352, 
corresponding to an epoch when the Universe was $\sim$1.8 Gyr old.
The analysis of its spectrum shows that this galaxy, hosting an AGN, formed and assembled 
 $\sim$2$\times$10$^{11}$ M$_\odot$ shaping its morphology within the $\sim$600 Myr
preceding the observations, since $z$$\sim$4.6.
The stellar population has a mean mass-weighted age 
of 400$^{+30}_{-70}$ Myr and it is formed  
between $\sim$600 Myr and $\sim$150 Myr before the observed epoch, this latter being the time since 
quenching.
Its high stellar velocity dispersion, $\sigma_e$=409$\pm$60 km s$^{-1}$, confirms
the high mass (M$_{dyn}$=$2.2(\pm0.4)$$\times$10$^{11}$ M$_\odot$) and the high
mass density 
($\Sigma_e^{M^*}$=$\Sigma_{1kpc}=3.2(\pm0.7)\times10^{10}$ M$_\odot$ kpc$^{-2}$), 
suggesting a fast dissipative process at its origin.
{The analysis points toward a supersolar metallicity, [Z/H]=0.25$^{+0.006}_{-0.10}$, 
in agreement with the above picture, suggesting a star formation 
efficiency much higher than the replenishment time.
However, sub-solar metallicity values cannot be firmly ruled out by our analysis.}
Quenching must have been extremely efficient to reduce the star formation to 
SFR$<$6.5 M$_\odot$ yr$^{-1}$ in less than 150 Myr. 
This could be explained by the presence of the AGN, even if a causal relation
cannot be established from the data.
C1-23152 has the same stellar and physical properties of the densest ETGs in the local 
Universe of comparable mass, suggesting that they are C1-23152-like galaxies 
which evolved to $z=0$ unperturbed. 
\end{abstract}



\section{Introduction} \label{sec:intro}
{ In the classical paradigm of hierarchical galaxy formation}, 
high-mass early-type galaxies (ETGs) grow their stellar mass mainly ex-situ 
through subsequent mergers of 
smaller pre-existing galaxies over timescales comparable to the Hubble time \citep[e.g.,][]{delucia06}. 
In seeming contrast, studies of local ETGs show that they follow tight 
scaling relations between stellar population properties (age and metallicity) and physical properties 
(stellar velocity dispersion, mass and size), 
whose tightness is not consistent with a predominant merging process \citep{nipoti12}
but requires a short star formation event \citep[e.g.,][]{renzini06}.
These studies also suggest that the larger the mass of a galaxy the higher the redshift at 
which most of its stars formed  
and the shorter the duration of star formation \citep{thomas10}.
Some simulations suggest that an early intense burst of star formation followed by 
a rapid quenching is required to reproduce the structural properties of ETGs and 
to match the tight scaling relations  \citep[e.g.][]{ciotti07, naab07,oser12,brooks16}

Local studies also show that the larger the mass of a galaxy, the higher the stellar metallicity
and the older the stars \citep[e.g.,][]{gallazzi06}.
This is rather counter intuitive since older stellar populations are expected with lower
metallicity than younger ones.
However, these properties 
result from the integrated effect of galaxy evolution across the whole Hubble time, making it
difficult to disentangle evolutionary from formation processes.
Since stellar metallicity is most sensitive to the efficiency of gas replenishment during
star formation and to the quenching mechanism \citep[e.g.][]{peng15,maiolino19},
the measurement of stellar metallicity in massive ETGs at high redshift would provide
important information on these two fundamental processes of galaxy formation. 

In the last decade, deep and wide photometric surveys
have discovered massive (M$^*$$>$10$^{11}$ M$_\odot$) quiescent galaxies
at z$>$2-3, i.e. formed during the first 2-3 Gyr 
of cosmic time \citep[e.g.][]{marchesini10,cassata13,straatman14}. 
Spectroscopic observations have confirmed the high redshift nature
for some of them, establishing the presence of passive galaxies at 2$<$$z$$<$3 with ages 
comparable to the age of the Universe at that redshift 
\citep[e.g.,][]{cimatti08,kriek09,gobat12,vandesande13,belli14,belli17}

More recently, spectroscopic observations have confirmed the presence
of massive passive galaxies at $z$$>$3 \citep{marsan15,glazebrook17,schreiber18,tanaka19, tanaka20,forrest20a,forrest20b,valentino20,deugenio20}.
The importance of pushing in the search for passive massive
galaxies to higher redshift is due to the constraints on models of galaxy
formation coming from the short 
cosmic time at the disposal  
of galaxies to assemble their stellar mass.
Indeed, the latest generation of galaxy formation models, 
e.g. Illustris \citep{wellons15} and TNG300 
\citep{springel18}, has difficulty in reproducing these galaxies 
since, at $z>3$,
there is little time ($<$2 Gyr) for smaller progenitors to build-up their stellar mass, halt
star formation and completely merge \citep{forrest20a}.

For some of these massive passive galaxies, the age of their stellar
populations has been constrained through 
a parametric fitting 
of their spectra and broad-band photometry, showing a large range in their ages \citep{glazebrook17,schreiber18,tanaka19,forrest20a,forrest20b,valentino20}.
So far, it has been possible to measure the stellar velocity dispersion for only one of these objects, dynamically confirming the high-mass of the system \citep{tanaka19}. 

While these massive passive galaxies at high redshift seem to have already completed 
their assembly, nothing is known about their stellar metallicity.
The unknown stellar metallicity does not allow 
for constraining either the timescale of the star formation or the possible 
quenching mechanisms, both affecting stellar metallicity differently \citep[e.g.][]{peng15}.
 Additionally, the uncertainty in measured stellar ages for some of them does not 
 allow for precise 
 estimates of their assembly time and the duration of the quenching process.
How ETGs have accreted their stellar mass, whether 
in-situ through a main star formation event in $<$1 Gyr 
\citep[e.g.,][]{thomas10}, or rather ex-situ through mergers in $>$2-3 Gyr \citep[e.g.,][]{boylan08},
or even through both processes at different times in a two-phase formation scenario \citep{oser10, hill17,newman18}, are still open and debated issues.

We have simultaneously measured stellar age, metallicity and velocity 
dispersion for C1-23152, a galaxy at $z=3.352$ \citep{marsan15},
when the Universe was less than 2 Gyr old,
with very deep (17 hours of integration) spectroscopic observations carried out at 
the Large Binocular 
Telescope (LBT).
In this paper we present the analysis and results.
The paper is organized as follows.
In \S\ 2 we summarize the properties of C1-23152 and describe 
the comparison data for ETGs at redshift $z$$\sim$0.
In \S\ 3 we describe the observations and the data reduction.
In \S\ 4 we present the spectrum and analyse its properties. 
In \S\ 5 we derive the stellar age and metallicity, and constrain the star formation
history; we derive the stellar mass and we measure stellar velocity dispersion. 
In \S\ 6 we discuss the results and present our conclusions.

Throughtout, we use the cosmological parameters H$_0$=70 km s$^{-1}$ Mpc$^{-1}$,
$\Omega_{\Lambda}$=0.7 and $\Omega_{M}$=0.3.
Magnitudes are given in the AB photometric system.
All the radii presented in the paper are circularized.
A \cite{chabrier03} initial mass function (IMF) is assumed throughout this paper.

\section{Data and Models}
\subsection{Galaxy C1-23152 at $z=3.35$}
\label{sec:data}
Galaxy C1-23152 (R.A. = 10$^h$00$^m$27$^s$.81, decl. = +02$^d$33$^m$ 49$^s$.3; J2000) 
was first presented in \cite{marchesini10} in a photometric study of a stellar-mass complete 
sample of galaxies at 3$<$z$<$4 using the NEWFIRM Medium-Band Survey. 
C1-23152 was then spectroscopically confirmed by \cite{marsan15}  to be at $z$=3.352 using a 
combination of Keck-NIRSPEC, VLT-Xshooter, and GTC-Osiris spectra. 
The SED modeling on the combined spectra and broad-/medium-band photometry resulted 
in an ongoing star formation rate $<$7 Msun/yr, and negligible dust extinction \citep{marsan15}.
From the analysis of the emission lines and the infrared SED, C1-23152 was found to harbor 
a powerful type-2 quasar (QSO), with bolometric luminosity of $\sim$10$^{46}$ erg s$^{-1}$, 
only mildly contaminating the stellar emission, 
with a lower limit to the stellar mass of 1.9$\times$10$^{11}$ M$_\odot$
(as taken from \cite{marsan15} after scaling to Chabrier IMF). 
Structural properties were derived from the analysis of 
HST ACS F814W and WFC F160W images, resulting in an effective radius R$_e\simeq$1 kpc and a 
Sersic index n$\simeq$4.4 in F160W \citep{marsan15}.

\subsection{Local comparison samples} 
As comparison samples of local ETGs the following data have been considered: 
a sample of ETGs selected from SPIDER \citep{labarbera10} with $\sigma_e$$>$150 km s$^{-1}$ 
to represent the whole population irrespective of their mass and size; 
the sample of ETGs with $\sigma_e>$350 km s$^{-1}$ studied by \cite{bernardi06}
representing the most massive ETGs in the local Universe; the sample of compact 
ETGs with high velocity dispersion studied by \cite{saulder15}, representing ETGs with 
high mass density and extreme structural and dynamical properties; the sample of compact galaxies 
identified by \cite{damjanov15}, representing the 
population of small, compact and dense galaxies missed by the current ground-based 
surveys.

\subsection{Stellar population models}
In this analysis, we adopted the EMILES simple stellar population (SSP) models 
\citep{vazdekis15}, based on BaSTI isochrones \citep{pietrinferni04} and Galaxy abundance ratios ([Fe/H]=[Z/H]) as a reference library, assuming a 
\cite{chabrier03} stellar initial mass function (IMF), 
with ages spanning [0.06; 2.0] Gyr, and 11 metallicity [Z/H] in the range [-2.32; 0.26] (a total of 162 SSPs).
{Notice that models with [Z/H]=0.4, although available, 
were not used in the analysis as they 
have lower quality than the other models \citep[see][for details]{vazdekis15}}. 
{ We also considered a set of MILES $\alpha$-enhanced models \citep{vazdekis15}
for which [$\alpha$/Fe]=0.4, spanning the same ranges of ages and metallicity of
the reference set.} 
In the optical spectral range, these models have a FWHM spectral resolution of 2.5 \AA\ \citep{beifiori11}, higher than the rest-frame 
resolution ($\sim$3 \AA) of the LBT-LUCI spectrum of C1-23152. 

The dependence of the stellar population properties on the adopted models
was tested by considering also BC03 SSPs \citep{bruzual03} with a Chabrier IMF, 
20 ages in the 
range [0.06; 2.0] Gyr and 5 metallicities in the range [-1.7; 0.4], and \cite{maraston11} 
MILES-based models (M11) with Chabrier IMF, including 20 ages in the range [0.06; 2.0] Gyr and 5 metallicities in the range [-2.3; 0.3] (see \S\ \ref{sec:ana}).

\section{Spectroscopic observations and data reduction}
Long slit spectroscopy of galaxy C1-23152 was obtained at the Large Binocular Telescope (LBT) 
with the two LBT Utility Cameras in the Infrared \citep[LUCI1 and LUCI2;][]{ageorges10} in twin (binocular) configuration, 
for a total effective integration time of 17.3 hours.
Observations were carried out on January 2018 and January 2019, with a seeing 
FWHM=0.8-1.0 arcsec (in the visual), with filter HK coupled with the 
grism G200 sampling the wavelength range 15000-23500 \AA\ covering the rest-frame range 3450 \AA$<\lambda_{rest}<$5300 \AA\ at the redshift of the galaxy, at 4.35 \AA/pix. 
We adopted a slit width of 0.75 arcsec resulting in a spectral resolution R$\simeq$1267(1650) in H(K) 
($\Delta\lambda$$\sim$13 \AA\ for both the bands).
Observations consisted of a sequence of exposures of 300 s each 
taken at dithered (ABBA) positions offset by $\sim$5 arcsec, summing up to 203 images.
A bright pivot star { (U0900.06569623; RA=0:00:32.39, Dec=+02:34:07.08; K$_{Vega}$=13.14 mag, spectral type M)} 71 arcsec away from the target galaxy was included in the slit to 
ensure accurate slit centering and alignment of dithered sequences, as well as
 to check the flux calibration and the correction for telluric absorption lines 
 in the data reduction phase. 

The 2D spectra were processed at the Italian LBT Spectroscopic
Reduction Center with a reduction pipeline optimized for LBT data \citep{scodeggio05,magrini12}.
For each one of the two observing runs (January 2018 and January
2019), calibration frames were created both for LUCI1 and LUCI2. A
bad pixel map was generated from flats and darks, while a ``master"
dark and a ``master" flat were created averaging a set of darks and
spectroscopic flats, respectively. 
Each spectral image was
independently corrected for cosmic rays and bad pixels, and then dark
and flat-field corrections were applied.

For the wavelength calibration,
an inverse solution of the dispersion was created for each observing day, and for
each spectrograph. The mean accuracy reached at the center of the H
and K bands is 0.5 \AA\ (rms).  For each frame, the slit was extracted and
wavelength calibrated, removing any curvature due to the optical
distortions.

Subtraction of emission from sky was performed following the method developed
by \cite{davies07} on the 2D-extracted and wavelength-calibrated spectra.
Further sky residuals along the spatial direction were removed by fitting
and subtracting the signal for each column of the spectrum.

{ Particular attention was paid to the relative flux calibration since it
can affect the shape of the spectrum.}
For each observing day, a sensitivity function, as well as telluric
absorption correction, was obtained for LUCI1 and LUCI2 using a
telluric star observed close in time and air-mass to the scientific
target. 
The wavelength/flux-calibrated and sky-subtracted spectra
obtained in the different nights were finally stacked
together.
Offsets between different frames were estimated using the
 pivot stars in the science exposures.

\begin{figure}
\centerline{
\includegraphics[width=8.5truecm]{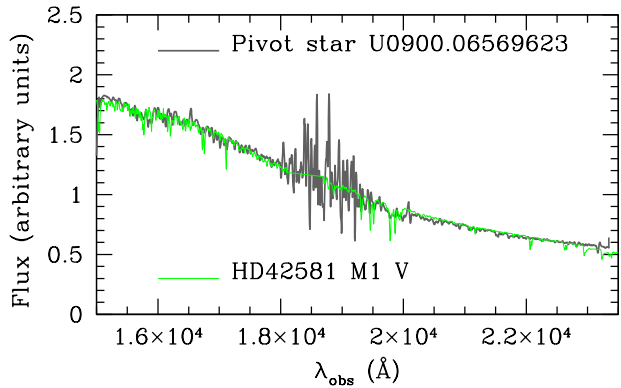}}
\caption{\label{fig:pivot} The final LBT-LUCI 1D spectrum of the pivot star
U0900.06569623 smoothed over 4 pixels ($\sim$17 \AA; gray curve) is compared with the spectrum of the main sequence M-type star HD42581 (green curve).
The two spectra have been normalized to the mean flux measured in the wavelength range
1.7-1.8 $\mu$m.
The comparison shows the perfect recovery of the true shape of the spectrum 
thanks to the careful correction for the response function of the instrument.}
\end{figure}

{ In Fig. \ref{fig:pivot}, the final LBT-LUCI 1D spectrum of the pivot star
U0900.06569623 (gray curve) is compared with the spectrum of the main sequence M1
spectroscopic standard star HD42581 (green curve)\footnote{The spectrum has been 
taken from the NASA Infrared Telescope Facility (IRTF) spectral library
http://irtfweb.ifa.hawaii.edu/~spex/IRTF\_Spectral\_Library/}.
The reliability of the continuum shape is a fundamental issue in the analysis
of a galaxy spectrum since its shape depends on the age and
metallicity of the underlying stellar population, and on the dust extinction.
This comparison shows a very good agreement, demonstrating the excellent recovery 
of the true shape of the continuum  and, therefore, the excellent 
quality of the relative flux calibration for C1-23152 spectrum.
}

In order to extract the 1D spectrum of the galaxy, we first removed the residual 
background component due to the bright star observed in the 
scientific exposures. 
This component was accurately removed by running the IRAF task \texttt{background} on the 
two-dimensional stacked spectrum. 
The center of the 2D galaxy spectrum was accurately modeled by fitting the 
profile along the spatial direction with a double Gaussian function, in wavelength bins, 
excluding pixels contaminated by sky residuals/telluric absorption, in each bin. 
The one-dimensional (1D) spectrum of the galaxy was extracted within a region of $\pm$ 3 pixels (0.75 arcsec) around 
the photometric center of the galaxy. 
We adopted an extraction radius of 3 pixels, as this turned out to maximize the S/N ratio 
of the extracted spectrum.

\section{The spectrum of C1-23152}
\begin{figure*}
\centerline{
\includegraphics[width=16.5truecm]{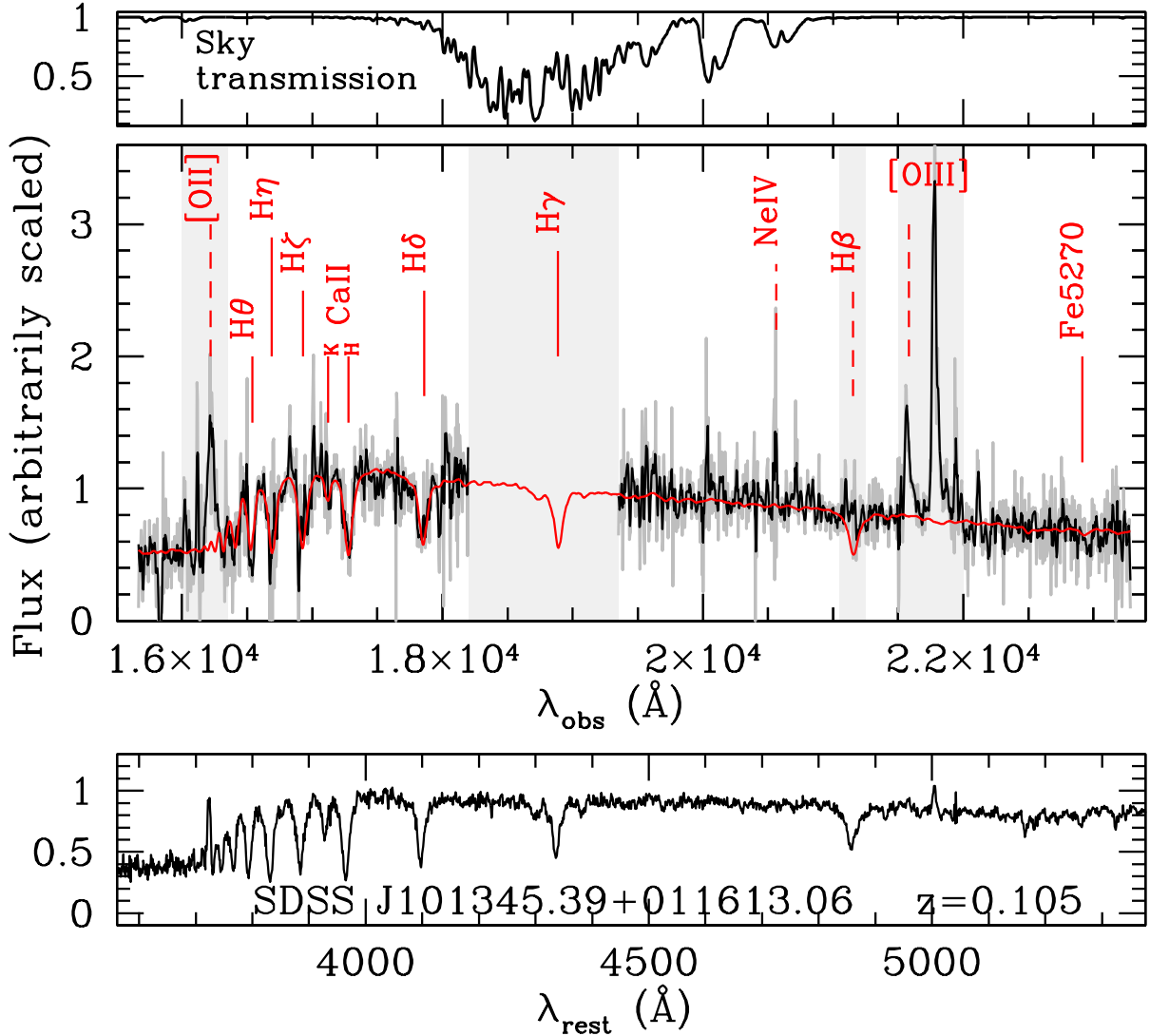}}
\caption{\label{fig:spec} Spectrum of galaxy C1-23152. 
The top panel shows the atmospheric transmission in the
wavelength range of observations. 
In the middle panel the one-dimensional spectrum of galaxy C1-23152
is shown in the original form (dark-gray curve, 4.35 \AA/pix, S/N$\simeq3$ \AA$^{-1}$) 
and smoothed by a boxcar filter over three pixels 
(black curve) corresponding to the instrumental resolution ($\Delta\lambda\simeq$13 \AA\ FWHM).
The main absorption and emission lines are marked by solid and dashed lines, respectively.
The red curve is the best-fitting composite model obtained with \texttt{STARLIGHT} (see \S\ \ref{sec:ana}).
The shaded gray regions are those masked in the fitting because of bad sky transmission
or the presence of emission lines.
For comparison, the bottom panel shows the observed spectrum of
a typical post-starburst galaxy in the local Universe selected from the Sloan Digital Sky Survey (SDSS).}
\end{figure*}

The 1D LBT-LUCI spectrum of C1-23152 is shown in Figure \ref{fig:spec},
middle panel.
The upper panel shows the atmospheric transmission in the wavelength range of observations.
The bottom panel shows, for comparison, the spectrum of
a post-starburst galaxy in the local Universe selected from the Sloan Digital Sky Survey (SDSS).
The spectrum of C1-23152 has a S/N$\simeq3$ \AA$^{-1}$ around 4000 \AA\
and it is characterized
by prominent emission lines and absorption features.

\subsection{Emission lines: Active Galactic Nucleus and star formation}
The spectrum shows strong [OII]($\lambda$3727) and [OIII]($\lambda\lambda$4958;5007) 
doublet emission lines and weak H$\beta$ emission.
The measured fluxes associated with these lines are
F$_{OII}$=4.0($\pm$0.5)$\times$10$^{-17}$ erg cm$^{-2}$ s$^{-1}$,
F$_{OIII}$=1.3$(\pm$0.3)$\times$10$^{-16}$  erg cm$^{-2}$ s$^{-1}$
and F$_{H\beta}$=4.3($\pm$2)$\times$10$^{-18}$ erg cm$^{-2}$ s$^{-1}$,
respectively\footnote{Fluxes were estimated by fitting a gaussian function to the line
after having removed the underlying continuum evaluated through a polynomial fitting of the 
regions adjacent to the line.}.
The high luminosity of [OIII]($\lambda$5007) line, L$_{OIII}$=1.3($\pm$0.3)$\times$10$^{43}$ erg s$^{-1}$
derived from the measured flux, cannot be produced by star formation and
suggests the presence of an Active Galactic Nucleus \citep[AGN;][]{francis91}.
This is also confirmed by the high value of the ratios [OIII]/H$_\beta\simeq$30 and
[OIII]/[OII]$>3$, with both of them in the AGN regime \citep{baldwin81} and by the 
detailed study of the infrared Spectral Energy Distribution (SED) presented by \cite{marsan15}.

The presence of an AGN makes it difficult to estimate the possible ongoing star formation
from the detected emission lines.\footnote{Using the [OII] emission, considering the relation 
[OII]/[OIII]=0.21 \citep{silverman09}, we obtained an average AGN contribution
L$(OII)_{AGN}$=2.7$(\pm0.2)\times$10$^{42}$ erg s$^{-1}$.
Therefore, the contribution of the SF to [OII] emission is 
L$(OII)_{SF}$=1.3$(\pm0.2)\times$10$^{42}$ erg s$^{-1}$ that, using the relation 
(scaled to Chabrier IMF) 
SFR=8.2$\times$10$^{-42}$L$_{OII}$ \citep{kennicutt98}, provides
 SFR$=$10 M$_\odot$ yr$^{-1}$.}
To circumvent this problem, 
we derived an upper limit to the current residual star formation rate (SFR)
assuming that all the H$\beta$ emission is due to star formation.
To this end, 
we corrected the measured flux for the expected H$\beta$ absorption 
(F$_{abs}$=6.7$\times$10$^{-19}$ erg cm$^{-2}$ s$^{-1}$) 
according to the best-fitting composite model (see \S\ \ref{sec:fitting}),
assuming no dust extinction.
We obtained
F$_{H\beta,corr}$=5.0($\pm$2.0)$\times$10$^{-18}$ erg cm$^{-2}$ s$^{-1}$.
Using the relation L$_{H\alpha}$=2.86L$_{H\beta}$ for case-B recombination \citep{moustakas05}, we obtained
L$_{H\alpha}$=1.4$(\pm0.6)\times$10$^{42}$ erg s$^{-1}$.
The corresponding upper limit to the star formation rate is SFR$=$6.6($\pm$3.0) M$_\odot$ yr$^{-1}$,
where we used the relation (scaled to Chabrier IMF) 
SFR=4.65$\times$10$^{-42}$L$_{H\alpha}$ \citep{kennicutt98}.
 This upper limit agrees well with the one derived from the fitting to the whole
SED by \cite{marsan15} (see above).

\subsection{\label{sec:abs} Absorption lines: age sensitive features}
\begin{figure}
\centerline{
\includegraphics[width=9truecm]{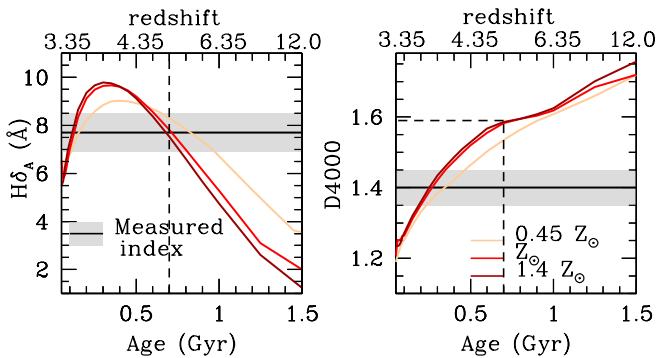}}
\caption{\label{fig:indices} Strength of the spectral indices 
H$\delta_A$ (left) and D4000 (right) expected for three stellar populations
of different metallicity (solid curves; orange curve, 0.45 Z$_\odot$;
red curve, Z$_\odot$ and brown curve 1.4 Z$_\odot$)
 as a function of age. 
Curves were obtained using EMILES SSPs \protect\citep{vazdekis15}. 
The black line represents the strength of the indices measured for C1-23152
(see Tab. \ref{tab:indices}), 
the gray shaded region represents the error at 68\% confidence level. 
The black dashed line marks the value that the two indices would have if
quenching would have taken place 700 Myr prior to observations, 
one of the two possible epochs suggested by the H$\delta_A$ index (left).
}
\end{figure}

The spectrum clearly shows the hydrogen Balmer absorption lines 
H$\theta$, H$\eta$, H$\zeta$, H$\delta$
along with CaII[H\&K], extremely weak Mgb and Fe($\lambda$5270).
Table \ref{tab:indices} summarizes the measured spectral indices (first row) 
and their errors at 1$\sigma$ (second row). 

\begin{deluxetable*}{cccccccccccc}
\caption{\label{tab:indices} Measured spectral indices for C1-23152}
\tablehead{
CN3883$^a$ &H$\theta$$^b$& H$\eta$$^b$& H$\zeta$$^b$ & H$\epsilon$$^b$& CaII[HK]$^c$ &  D4000$^d$ & D$_n$$^e$& 
H$\delta_A$$^f$ & H$\delta_F$$^f$ & Mgb$^f$ & Fe5270$^f$ }
\startdata
-0.07 & 5.8 & 7.8 & 9.9& 8.4& 7.0  &	1.40 & 1.09 & 7.7 & 5.5  & 0.5 & 5.6 \\
(0.04) & (1.0)& (1.3)& (1.1)& (0.9)&(1.6)  &(0.05) & (0.06) & (0.8) & (0.9) & (0.8) & (2.3) \\
\enddata
\tablecomments{Indices are corrected for galaxy velocity dispersion. The correction was derived by comparing the indices measured on the best-fitting model smoothed to the $\sigma_{obs}$ of the
galaxy, and those of the same model at the nominal resolution of the spectral library.
$^a$As defined in \cite{davidge94}; $^b$defined ad-hoc in this work to avoid 
as much as possible sky residuals in the pseudo-continuum regions; $^c$as defined by \cite{serven05}; $^d$as defined by \cite{bruzual83}; $^e$as defined by \cite{balogh99}; $^f$as defined by \cite{worthey97} and \cite{trager98}. 
Note that the coincidence with a strong sky emission line
makes the measured strength of Fe5270 rather unreliable.
For this reason, this feature was not considered in the absorption line fitting.}
\end{deluxetable*}

The Balmer lines are typical of the post-starburst phase since they are associated with hot, 
high-mass (1.5-2 M$_\odot$) rapidly evolving stars whose main sequence lifetimes are less than 800 
million years (Myr) \cite[e.g.][]{poggianti97}.
Their strengths are related to the elapsed time since 
the end of the last burst of star formation, i.e. the time since quenching.
Figure \ref{fig:indices} (left panel) shows the expected H$\delta_A$ index for a simple stellar population (SSP, EMILES models) seen at 
different ages for three different values of stellar metallicity (colored curves).
The figure shows that the strength measured for galaxy C1-23152, H$\delta_A=7.7\pm0.8$ \AA\ (black solid line), 
defines two possible quenching epochs, at $<$200 Myr and at $\sim$600-800 Myr,
for solar or supersolar metallicity.

The right panel of Fig. \ref{fig:indices} shows, instead, the expected amplitude of the 
D4000 index \citep{bruzual83}.
This discontinuity is produced by the opacity of the stellar atmospheres that, in turn,
depends on the ionized metals hence, on the stellar temperature.
Hot stars, responsible for H$\delta$, do not contribute to the amplitude of D4000
since their elements are multiply ionized and their opacity is low.
Therefore, D4000 index, contrary to H$\delta$, is dominated by low mass stars
and, hence, its amplitude is strictly related to the age of the bulk of the stellar population.
The measured D4000=1.40$\pm$0.05 (black solid line, right panel of Figure \ref{fig:indices})
suggests a mean age for the stellar population in the range 250-450 Myr.
This mean age rules out the solution for quenching to be 600-800 Myr before observations 
(which would require D4000$>$1.6, black dashed line,  
$>$3$\sigma$ in excess of the measured value), and constrains
the bulk of the star formation to occur within the past $<$700 Myr.
Combined with the H$\delta_A$ measurement, the time since quenching
is restricted to $<$200 Myr.
We verify these conclusions following a more careful and detailed analysis in \S\ \ref{sec:ana}. 

\subsection{Absorption lines: metallicity sensitive features}
\begin{figure}
\centerline{
\includegraphics[width=9truecm]{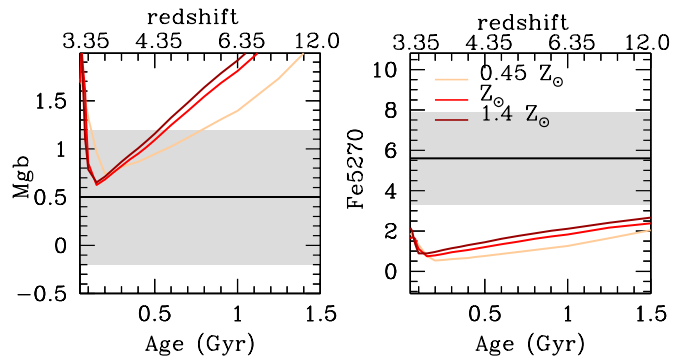}}
\caption{\label{fig:mgb} Strength of the spectral indices
Mgb($\lambda$5175) (left) and Fe($\lambda$5270) (right) expected for a stellar population
with different metallicity seen at different ages (colored curves), and
measured for galaxy C1-23152 (black line).
Symbols are as in Fig. \ref{fig:indices}.
}
\end{figure}

\begin{figure}
 \centerline{
 \includegraphics[width=8.5truecm]{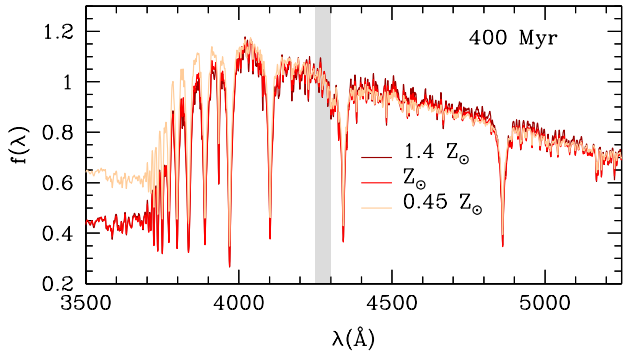}}
\caption{\label{fig:emiles} 
The colored curves represent three EMILES SSPs 400 Myr old for
three metallicity values, 0.45 Z$_\odot$ (orange), Z$_\odot$ (red)
and 1.4 Z$_\odot$ (dark red). 
The gray region marks the interval 4250-4300\AA\ where the SSPs are normalized.
}
\end{figure}
Three main metal lines fall in the LUCI spectrum of  C1-23152: 
CN3883, Mgb($\lambda$5175) and Fe($\lambda$5270), all of them having low S/N 
(see Tab. \ref{tab:indices}).
The coincidence of Fe($\lambda$5270) feature with a strong sky emission 
line makes the measured strength of this feature rather uncertain.

{Figure \ref{fig:mgb} shows the line-strength of Mgb($\lambda$5175)
(left) and Fe($\lambda$5270) (right) for EMILES SSPs with three different
metallicities (see colored curves) as a function of age.
The horizontal black lines are the line-strengths as measured for C1-23152.
These indices are widely used to derive stellar metallicity and abundances.
Models show that the two features are extremely weak and, as expected, 
weakly dependent on metallicity at young ages ($\leq$500 Myr).
Namely, for metallicity in the range [0.45-1.4]Z$_\odot$, the maximum variation of Mgb is 
less than 10\% for ages younger than 500 Myr and $<$30\% at 700 Myr.
A similar behavior is seen for Fe($\lambda$5270), with the difference that for ages older than 500
Myr the maximum variation is nearly constant, by about 25\%.
The weak sensitivity  of these indices to variations in metallicity at young ages 
is model independent, as we verified considering BC03 \citep{bruzual03} and \cite{maraston11}
(M11) models.
Therefore, given the young mean age of the stellar population these indices are 
not expected to provide significant constraints on stellar metallicity.
}

Figure \ref{fig:emiles} shows the spectra of three SSPs of different metallicity at
fixed age, 400 Myr, in the wavelength range 3500-5200 \AA.
These models show that, absorption features at shorter wavelengths and the continuum shape around D4000 are more dependent on metallicity, contrary to the features at longer wavelengths.
Therefore, at young ages, Balmer lines, D4000, and continuum shape can constrain 
stellar metallicity more than metal lines.
{ We will take advantage of this in the next section, being certain of the reliability 
of the continuum shape of our spectra (see \S\ 3 and Fig. \ref{fig:pivot})}.

\section{\label{sec:ana} Analysis}
In this section we describe the analysis performed to estimate the stellar age, metallicity and
velocity dispersion of C1-23152.
The aim is to constrain the formation timescale of this ETG, i.e. the time needed to fully assemble 
and shape this galaxy as seen at z=3.35, and the mechanism of mass growth, i.e. whether the stellar
mass of this galaxy was assembled in-situ through star formation or ex-situ through accretion.

\subsection{\label{sec:fitting} Age and metallicity estimates}
Stellar age and metallicity were derived through two different
methods: absorption lines fitting (ALF) and full spectral fitting (FSF).
\paragraph{ALF}\
{ We performed absorption line fitting \citep{labarbera13,saracco19} by comparing
the measured line-strengths with those predicted by SSP models of varying age and metallicity}.
The fitting was performed by minimizing the expression
\begin{equation}
 \chi^2(age,[Z/H])=\sum_j{{(O_j-M_j)^2}\over{s_j^2}}
\end{equation}
where the index $j$ runs over the selected set of spectral indices in the rest-frame 
range 3500-5200 \AA.
{ The main spectral indices considered in the fitting, with the exception of Fe5270, 
are summarized in Table \ref{tab:indices}.}
$O_j$ and $M_j$ are observed and model index values, where the latters depend on age and [Z/H], while
the $s_j$'s are the uncertainties on the observed indices.
Some mild extrapolation of indices, to up [Z/H]=0.3, has been performed following the same 
approach as in \cite{labarbera13}.
The resulting best-fitting light-weighted { SSP equivalent} age and metallicity values are 
age$_L$=0.24$^{+0.11}_{-0.05}$ Gyr and
[Z/H]$_L$=0.27$^{+0.03}_{-0.20}$, respectively.
The quoted errors were obtained by running the fitting procedure on a set of 100 simulated
spectra (see Appendix \ref{sec:sim}).

In Figure \ref{fig:alf} the Balmer line indices measured on the spectrum of C1-23152 are compared
to those expected from models at different ages and metallicity values.
H$\zeta$, H$\eta$, H$\theta$ and H$\iota$ all favour high metallicity values, consistent with D4000.
H$\gamma_F$ has a very large error bar, while H$\epsilon$ and H$\delta_A$ tend to favour sub-solar metallicity.
The net effect is that high-metallicity values are favored from the fitting of line-strengths, 
consistent with results from STARLIGHT (see below). 
Also, one should bear in mind that H$\delta$ and H$\epsilon$ may be 
affected by some emission contamination, as expected from the residual star formation 
(in particular H$\delta$), reinforcing the conclusion above.
{ The horizontal gray arrow in the H$\delta$ panel of Figure \ref{fig:alf}
shows the value of the index corrected by 15\%.
This is the median correction for emission filling to H$\delta$ absorption
derived empirically by \cite{goto03a} from a sample of 3300 local post-starburst
galaxies.\footnote{ Other empirical methods, as the one based on the D4000,
$\Delta$EW(H$\delta$)=-5.5D4000+11.5, provide larger corrections
\citep[30\%-40\%, e.g.,][]{miller02}}
It is reasonable to expect for H$\epsilon$ a correction not larger than the one
for H$\delta$.}

\begin{figure*}
 \centerline{
 \includegraphics[width=15truecm, height=12truecm]{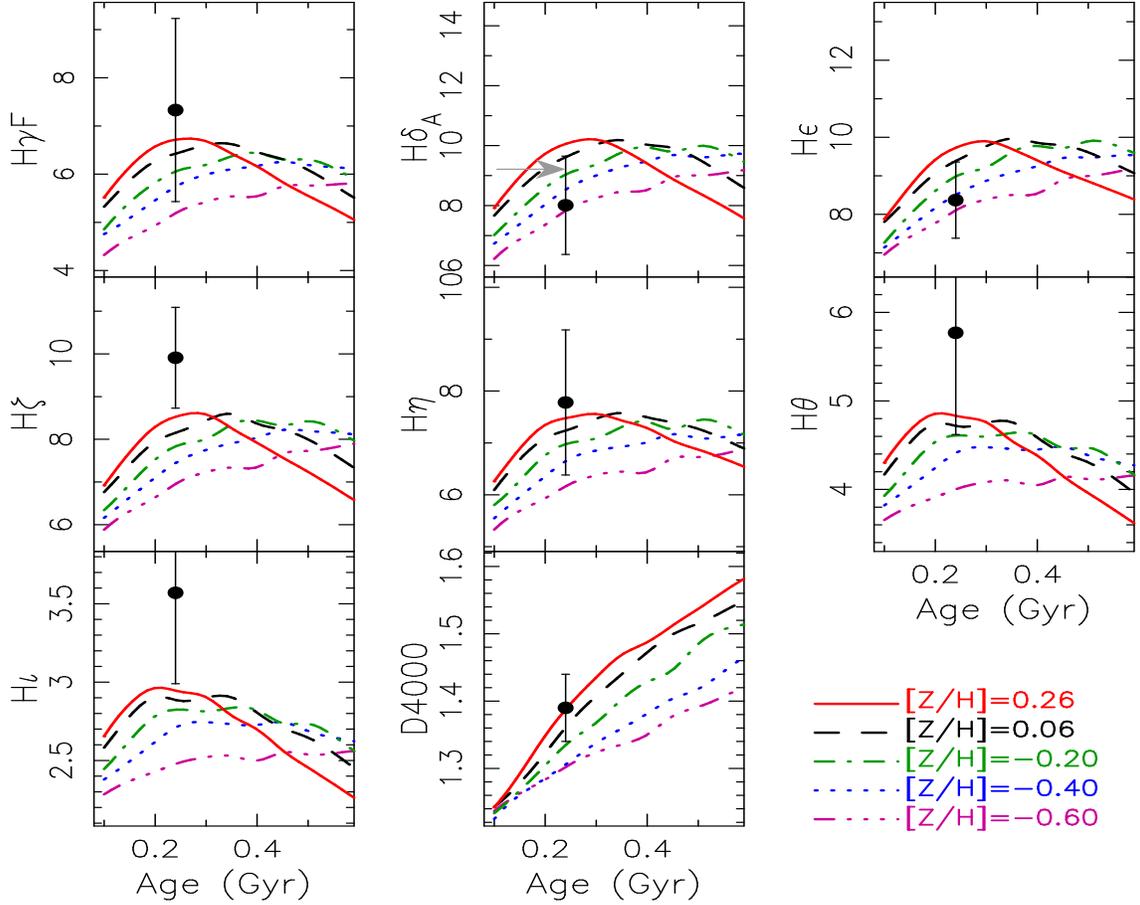}}
\caption{\label{fig:alf} C1-23152 absorption lines strength.
Balmer absorption lines indices measured on the spectrum of C1-23152
(black filled point) compared to those expected at different ages 
from EMILES SSPs with different metallicity values (coloured curves).
{ The horizontal gray arrow in the H$\delta$ panel shows the value of
the index corrected by 15\%\ to account for the average emission filling,
as derived by \protect\cite{goto03a} for local post-starburst galaxies (see also text).}
}
\end{figure*}

\begin{deluxetable*}{ccccccc}
\caption{\label{tab:fit} Stellar population properties}
\tablehead{
Age$_L$ & [Z/H]$_L$ & Age$_{M*}$ & [Z/H]$_{M*}$ & A$_V$&IMF& Method\\
(Gyr) & & (Gyr) & & (mag) & }
\startdata
0.24$^{+0.11}_{-0.05}$ & 0.27$^{+0.03}_{-0.20}$ &... &... &...& Cha & ALF\\
0.27$^{+0.05}_{-0.02}$ &   0.25$^{+0.007}_{-0.11}$   &   0.40$^{+0.03}_{-0.07}$   &   0.25$^{+0.006}_{-0.10}$ & 0.05$^{+0.06}_{-0.05}$& Cha&FSF\\
0.28$^{+0.05}_{-0.02}$ & [0.0]$^a$ & 0.41$^{+0.03}_{-0.07}$ & [0.0]$^a$ & 0.09$^{+0.05}_{-0.05}$ & Cha & FLF$^a$ \\
0.27$^{+0.05}_{-0.04}$ &   0.25$^{+0.00}_{-0.04}$  &	0.42$^{+0.06}_{-0.09}$   &   0.25$^{+0.00}_{-0.03}$ & 0.05$^{+0.12}_{-0.11}$& Sal&FSF\\
0.23$^{+0.05}_{-0.02}$ &   0.12$^{+0.10}_{-0.08}$   &   0.30$^{+0.05}_{-0.07}$   &   0.13$^{+0.10}_{-0.08}$ & 0.19$^{+0.10}_{-0.11}$& Cha&FSF+$\alpha$-enhanced\\
\enddata
\tablecomments{$^a$This fitting has been obtained at fixed solar metallicity. Errors are assumed
as those obtained with free metallicity.}
\end{deluxetable*}

\paragraph{FSF}\ 
In the full spectral fitting approach, stellar population
models are matched to the observed spectrum in wavelength (rather than
spectral index) space.  
To perform FSF, we have adopted the software \texttt{STARLIGHT} \citep{cid07, mateus07}, 
that fits a linear combination of SSP models (the ``base''), with different ages and
metallicities, to derive the best-fitting composite model to the
observed spectrum.  
The main advantage of this approach is that it is
non-parametric, i.e.  no a-priori assumption on the functional form
of the galaxy star formation history (SFH) is done.  
In this way, one can detect, in principle, multiple stellar components if present, and
constrain their metallicity.  
On the contrary, assuming a given functional form for the SFH, would necessarily imply 
a trade off between age, duration of star-formation, and metallicity. 
Of course, any non-parametric approach requires the robustness of the
detected stellar population components to be assessed (i.e. whether they are real, or
just result from some fluctuations in the observed spectrum), as we
extensively test here using Monte-Carlo simulations.

\texttt{STARLIGHT}  derives the best-fitting linear combination of SSPs using a
Markov Chain Monte Carlo (MCMC) algorithm.  
The model spectrum is the superposition of SSPs, with age Age$_i$ and metallicity Z$_i$, each one contributing with a different fraction ($x_i$) to the light and a different fraction ($m_i$) of 
stellar mass, taken from a pre-defined set of base spectra (162 EMILES SSPs in our case).
{There is no restriction to the number of spectral component (SSPs) that enters in the composite 
model.}
Light-weighted ($L$) and mass-weighted ($M$) age Age$_{L,M}$ and metallicity [Z/H]$_{L,M}$ 
can be defined
according to the relations \citep[e.g.,][]{asari07}
\begin{equation} 
{\rm Age}_{L,M}=\sum_i (x_i;m_i) {\rm Age}_i
\end{equation}
 and
 \begin{equation}
{\rm [Z/H]}_{L,M}=log \sum_i (x_i;m_i)Z_i/Z_\odot.
\end{equation}
Internal reddening in the range 0-2 mag was allowed in the fitting by considering both the Cardelli (CCM) 
\citep{cardelli89} and the Calzetti (HZ5) \citep{calzetti00} extinction laws,
with no difference in the results.
The spectral regions affected by bad sky transmission and by the presence of emission
lines were masked in the fitting.
The best fitting composite model, sum of  5 SSPs (see \S\ 5.2 for details), is shown in 
Figure \ref{fig:spec} as red curve.
The corresponding mean mass-weighted stellar age (eq. 2) is Age$_M$=400$^{+30}_{-70}$ Myr
(similar to that derived in \S\ \ref{sec:abs}),
and metallicity (eq. 3) is [Z/H]$_M$=0.25$^{+0.006}_{-0.10}$ (second row of Tab. \ref{tab:fit})\footnote{To quantify the maximum possible influence of the AGN on age and metallicity estimation, 
we subtracted the composite quasar spectrum by \cite{francis91} to the spectrum of C1-23152 and 
we run Starlight.
The QSO specrum was normalized to the rest-frame UV flux of the galaxy at $\sim$1400 \AA\
(filters IA624, IA679) assuming that all the observed flux was due to QSO.
We obtained Age$_M$=320$\pm$50 Myr and  [Z/H]=0.20$\pm$0.06, respectively.}, 
while the 
luminosity-weighted age and metallicity are Age$_L$=270$^{+50}_{-20}$ Myr and [Z/H]$_L$=0.25$^{+0.007}_{-0.11}$, respectively.
Notice the excellent agreement between the light-weighted age and metallicity values
obtained with two independent methods, ALF and FSF.
Errors (68\% confidence level) and stability of best fitting results  
were assessed by running \texttt{STARLIGHT} on a set of simulated spectra 
(see Appendix \ref{sec:sim}).

{ We checked the minimum $\chi^2$ solution found by \texttt{STARLIGHT}
by repeating the fitting with penalized PiXel-Fitting method,
\texttt{pPXF} \citep{cappellari04,cappellari17}, that allows for regularization 
\citep[see][for a comprehensive discussion of regularization in full spectral fitting]{cappellari17}.
The resulting best-fitting light-weighted age and metallicity values obtained
for four different degrees of regularization (parameter REGUL=1, 2, 3, 4) are
Age$_L$(Myr)=[254, 243, 213, 200] and [M/H]$_L$=[0.20, 0.20, 0.21, 0.20],
in agreement with the results obtained with other methods (see Tab.\ref{tab:fit})
} 

{ The degeneracy between age, metallicity and dust has been probed by repeating 
the fitting with \texttt{STARLIGHT} at fixed metallicity for the 11
different metallicities of the EMILES library.
We made use of the F-test to compare the $\chi^2$ of the 11 fits
with the best-fitting composite model.
We found that, fits down to metallicity [Z/H]=-0.60 are still within one sigma
from the best-fitting one.
For this metallicity, [Z/H]=-0.60, the corresponding light- and mass-weighted ages are  
Age$_L$=420 Myr and Age$_M$=480 Myr with an extinction A$_V$=0.01 mag.
In Table \ref{tab:fit}, the values obtained at fixed solar metallicity are reported 
for reference purposes.}

To test the effect of varying the IMF on age and metallicity estimates\footnote{The effect of 
a time-dependent IMF and of metallicity on age and star formation timescale is discussed 
in \cite{jerabkova18} and \cite{yan19}.}, we considered also a 
base of SSPs with Salpeter \citep{salpeter55} IMF.
This IMF provides slightly (not significantly) older age than Chabrier IMF,
as reported in Table \ref{tab:fit}. 

{ In the case of $\alpha$-enhanced MILES SSPs (see \S\ 2), the best
fitting composite model provided a mass-weighted stellar age Age$_M$=300$\pm$50 Myr and
metallicity [Z/H]$_M$=0.13$^{+0.10}_{-0.08}$, while the luminosity-weighted values are
Age$_L$=230$\pm$50 Myr and [Z/H]$_M$=0.12$^{+0.10}_{-0.08}$}.

Finally,
Table \ref{tab:fitmod} lists the best fitting age and metallicity values obtained with 
BC03 \citep{bruzual03} and M11 \citep{maraston11} models.
The results are consistent with the reference EMILES library, with the exception
of the light-weighted age obtained with M11 library, older than the others.

{ It is worth noting that metallicity, even if affected by larger uncertainty 
than age, is found to be always higher than solar ([Z/H]$>0$) independent of the models 
and the methods (see Tab. \ref{tab:fit} and \ref{tab:fitmod}).
Moreover, the best-fitting metallicity values obtained for the 100 simulated spectra
are all higher than solar (see Fig. \ref{fig:simstar}), i.e. lower metallicity
values never provide better fit, regardless the age and the extinction.
This suggests that, actually, age, metallicity and dust extinction 
are not completely degenerate with respect to
absorption features and continuum shape.

Even if acceptable fit can be obtained for sub-solar metallicity values, all the 
results point toward a supersolar metallicity for C1-23152, 
regardless of the method, the models and the IMF.}

\begin{deluxetable}{ccccc}
\caption{\label{tab:fitmod}Stellar population properties for diffent library of models}
\tablehead{
Age$_L$ & [Z/H]$_L$ & Age$_{M*}$ & [Z/H]$_{M*}$ & Model\\
(Gyr) & & (Gyr) & & }
\startdata
0.27$^{+0.05}_{-0.02}$ &   0.06$^{+0.10}_{-0.05}$   &   0.30$^{+0.07}_{-0.05}$   &   0.10$^{+0.08}_{-0.05}$ & BC03\\
0.80$^{+0.01}_{-0.10}$ &   0.21$^{+0.00}_{-0.10}$  & ...  &  ...  & M11\\
\enddata
\end{deluxetable}

\subsection{Stellar populations and SFH}
\begin{figure*}
\includegraphics[width=12truecm]{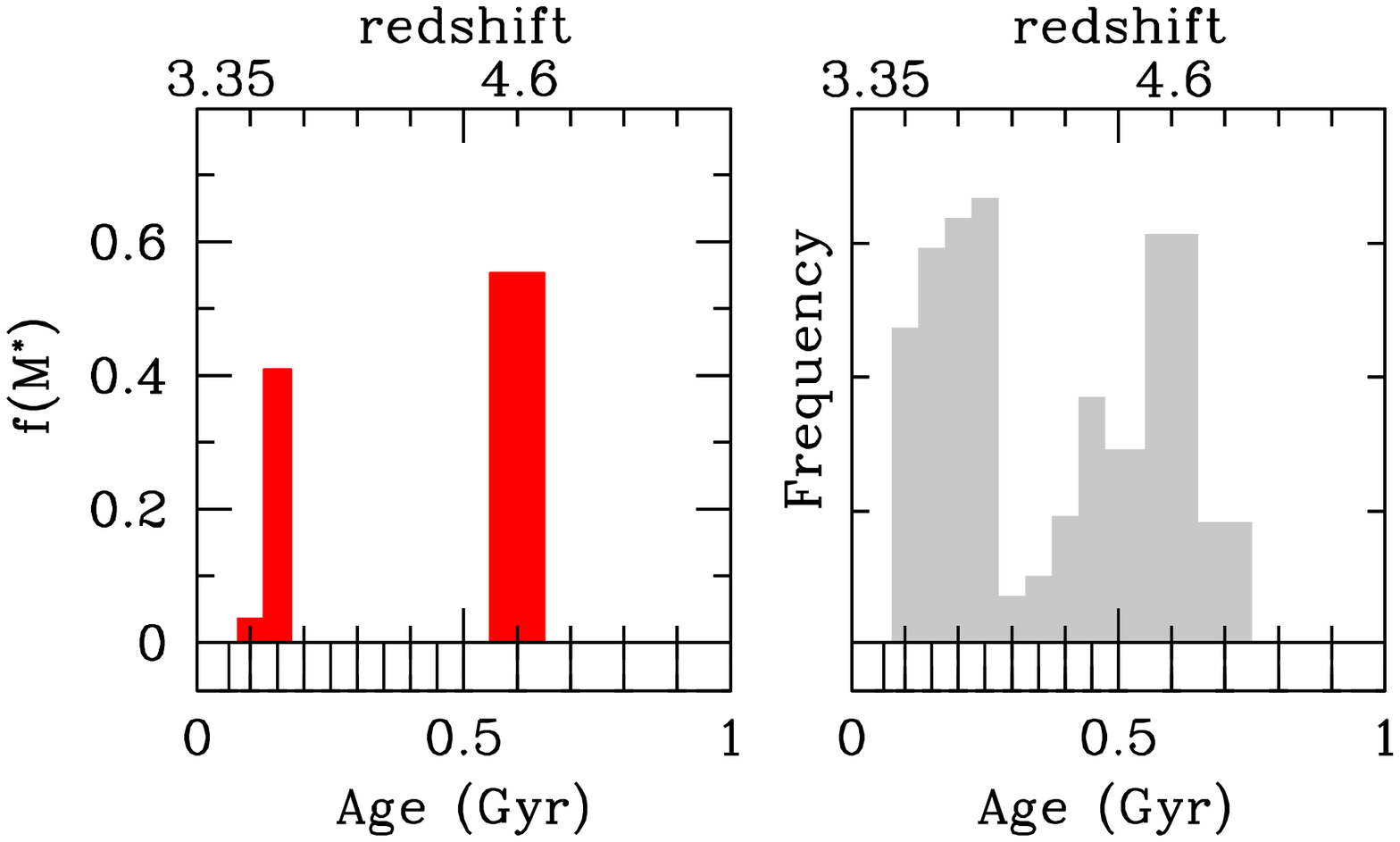}
\includegraphics[width=6.6truecm]{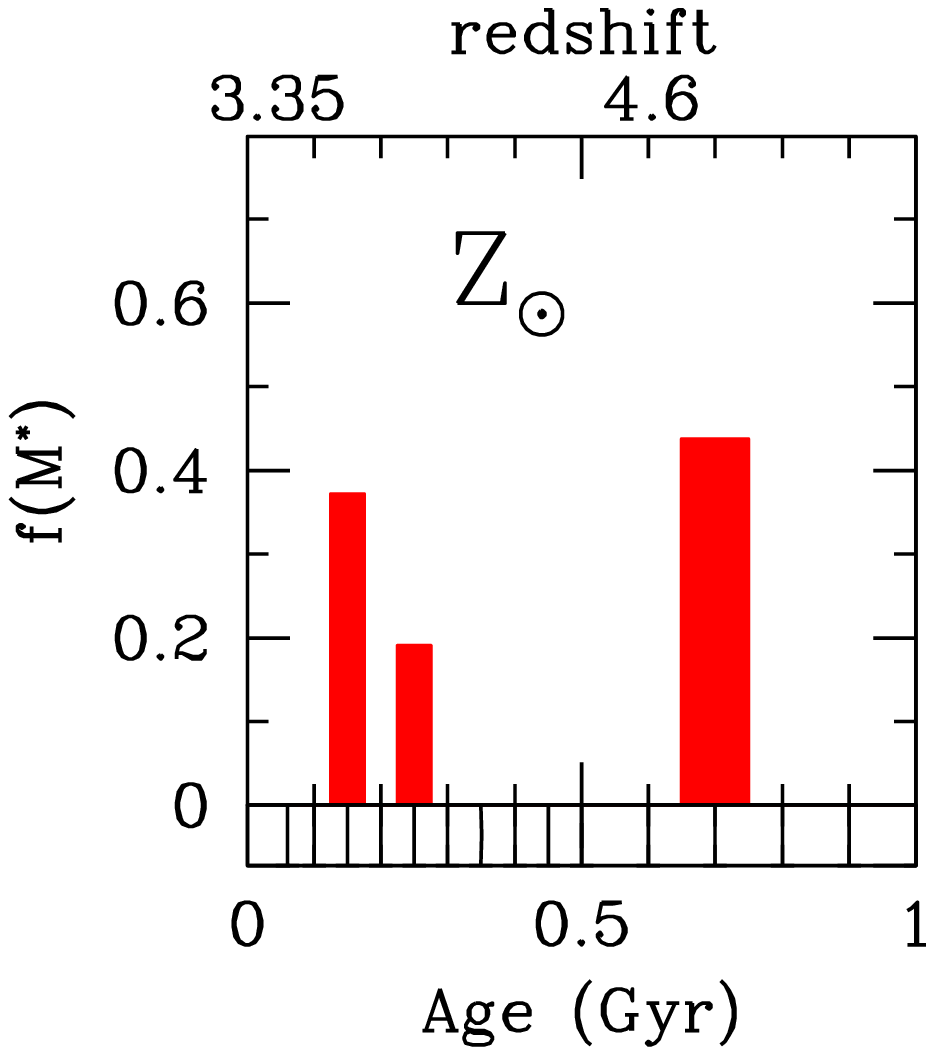}
\caption{\label{fig:sfh}Stellar populations and star formation history. 
The red histogram on the left panel shows the relative fraction of stellar mass
associated to the SSPs that compose the best-fitting composite model of the spectrum of C1-23152 obtained
 with \texttt{STARLIGHT} 
(see \S\ \ref{sec:ana} and second row of Tab.\ref{tab:fit}). 
The middle panel shows the distribution (gray histogram) of the best-fitting SSP components 
obtained by running \texttt{STARLIGHT} on the 100 simulated spectra
(see Appendix \ref{sec:sim}).
The right panel is the same as the left panel but for fixed solar metallicity 
(see third row of Tab. \ref{tab:fit}).
The bar-code on the bottom of the figures shows the age grid of the base of SSPs.
}
\end{figure*}
\begin{figure}
\includegraphics[height=2.5truecm, width=8.truecm]{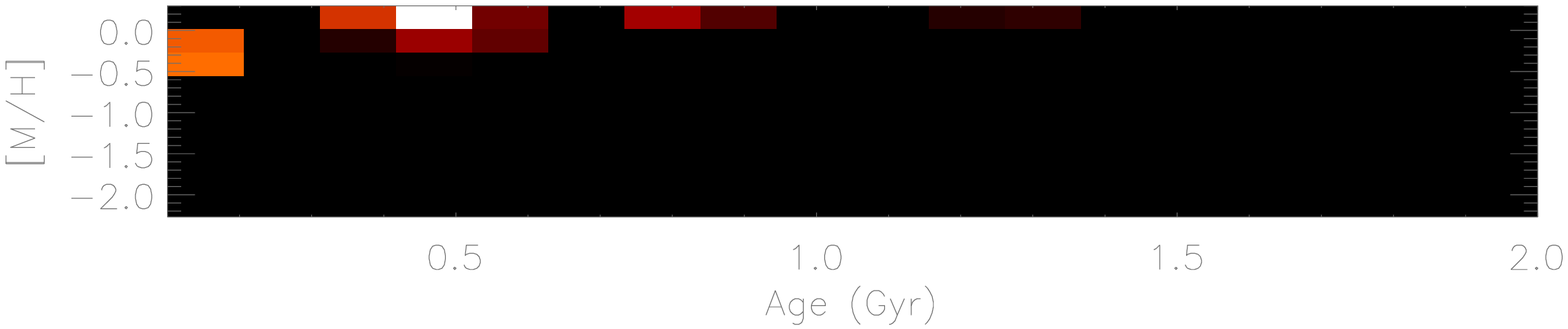}
\includegraphics[height=2.5truecm, width=8.truecm]{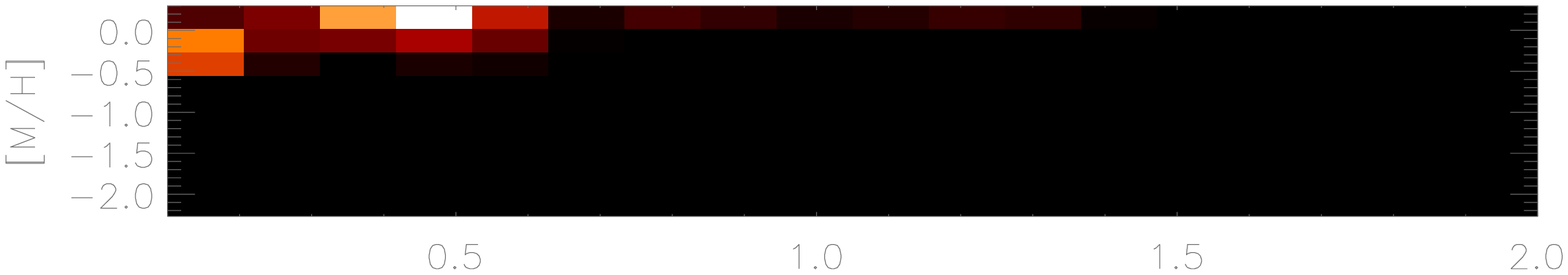}
\includegraphics[height=2.5truecm, width=8.truecm]{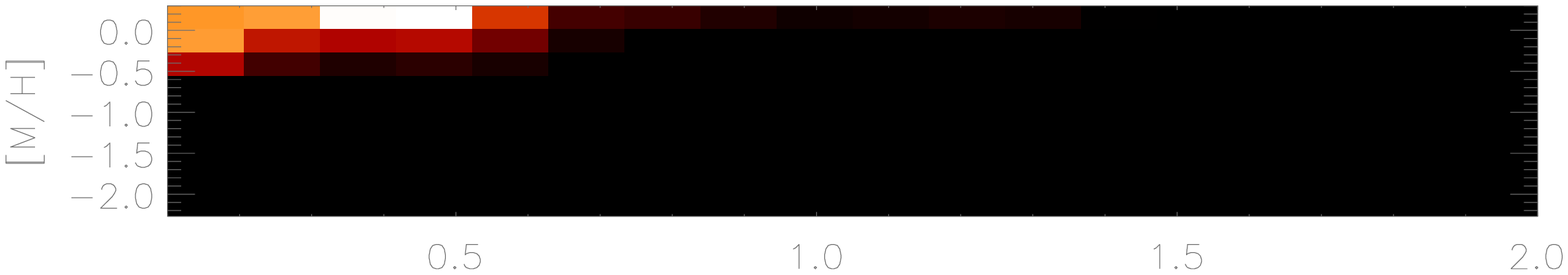}
\includegraphics[height=2.8truecm, width=8.truecm]{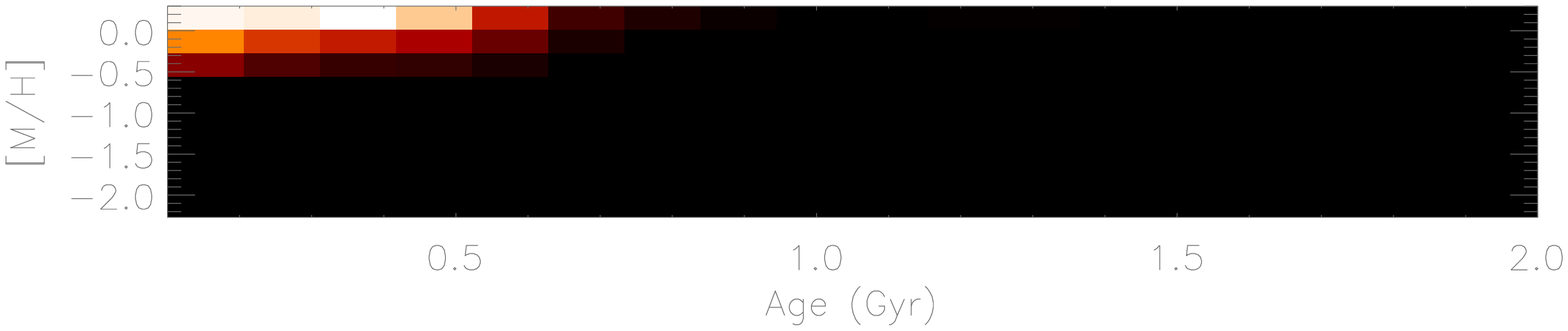}
\caption{\label{fig:ppxf} Weighted maps of the \texttt{pPXF} regularized solutions.
Each panel shows the contribution of the SSPs composing the best-fitting composite
model in the different age and metallicity intervals considered. 
The four maps have been obtained for four different values of regularization, 
(from top to bottom) REGUL=[1, 2, 3, 4].
The corresponding light-weighted age and metallicity values are:
Age$_L$(Myr)=[254, 243, 213, 200] and [M/H]$_L$=[0.20, 0.20, 0.21, 0.20],
in agreement with the results obtained with the other methods (see Tab. \ref{tab:fit}).
The SSPs contributing to the best-fitting model
are all in the age range 0.1-0.6 Gyr, confirming the interval of the star
formation  derived with
\texttt{STARLIGHT} and roughly constrained in \S\ 4.2.
}
\end{figure}

Figure \ref{fig:sfh} shows the contribution to the stellar mass of the different SSPs that compose the
best-fitting composite model.
\texttt{STARLIGHT} fitting detects two main stellar components (out of the five)
contributing for more than $\sim$95\% of the stellar mass:
an older one peaked about 600 Myr prior to the epoch of observation, setting the initial formation redshift to $z_f\simeq4.6$, and a younger one peaked 150 Myr prior to observation, constraining the redshift of
quenching  at $z\simeq3.6$.
The presence of two main stellar components is also suggested by the distribution 
of the best-fitting components obtained on the 100 simulated spectra shown in the middle panel of Figure \ref{fig:sfh}.
{ The fitting at fixed metallicity described in \S\ 5.1 does not 
significantly affect the SFH in the sense that, the star formation is always 
constrained within an interval $\Delta t_{SF}$$\simeq$450-500 Myr with an older 
main component and one or more younger components.
As example, in the right panel of Figure \ref{fig:sfh}, it is shown the SFH at 
solar metallicity.

We probed the possible degeneracy in the solutions, by repeating the
full spectral fitting with pPXF (see \S\ 5.1).
Fig. \ref{fig:ppxf} shows the weighted age-metallicity map for the SSPs 
contributing to the best-fitting model obtained for each of the four regularization 
values (REGUL=1, 2, 3, 4).
In all the cases, the SFH is constrained within the interval 0.1-0.6 Gyr,
with an increasing smoothness for increasing values of regularization. 

These results demonstrate that the stellar population is not coeval as 
resulting from a SFH extended over an interval 450-500 Myr, possibly characterized either by two main massive episods or even by a more continuous distribution.}

\begin{figure}
\includegraphics[width=8.5truecm]{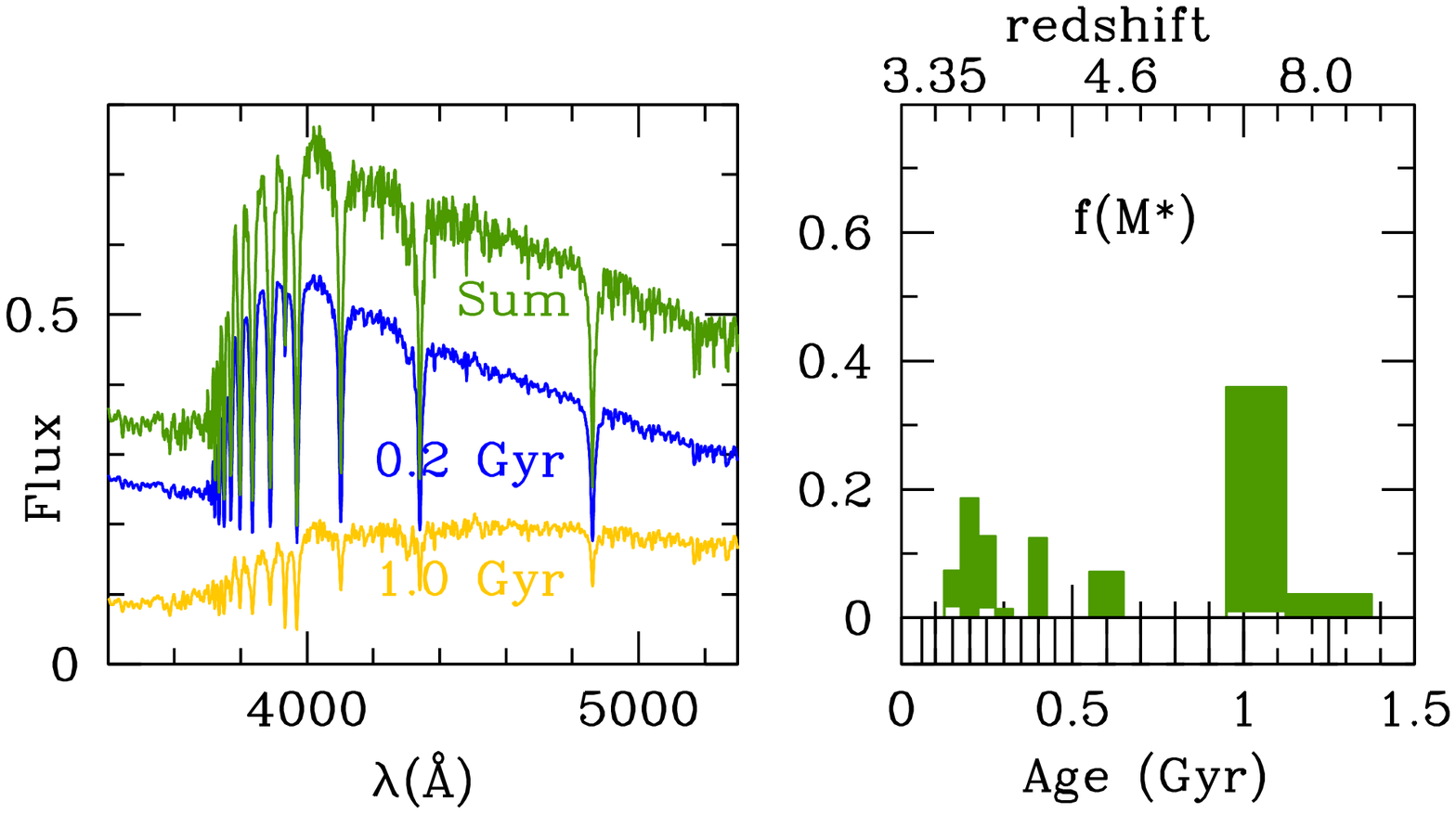}
\caption{\label{fig:mara} Left - Simulated spectrum (green curve) sum of a SSP 0.2 Gyr old
(blue curve)
representing 35\% of the stellar mass (75\% of the flux at 4000 \AA), and
a SSP 1.0 Gyr old (yellow curve) accounting for the remaining 65\% (25\%).
Right - Relative fraction of stellar mass associated to the SSPs that compose 
the best-fitting composite model to the simulated spectrum.
The bar-code on the bottom shows the age grid of the base of SSPs.
}
\end{figure}

{It is interesting to note, that the spectra of quiescent 
galaxies at $z$$>$3 reported to date appear in a post-starburst phase \citep[see e.g.,][]{deugenio20},
as C1-23152.
As such, the spectra are sensitive to the time since quenching, 
being dominated by the youngest stellar population (i.e. the stars
formed around quenching) which outshine older stars.
This effect could make it difficult to detect stellar populations formed before
quenching, i.e. to infer the actual time over which galaxies formed stars and the age 
of the bulk of the galaxy stellar population.

We tested the ability of spectral fitting in detecting  stellar populations older
than the outshining youngest one, by running \texttt{STARLIGHT} 
with the same setup used for the spectrum of C1-23152,
on a synthetic spectrum composed of a SSP 200 Myr old accounting for 35\% (75\%) 
of stellar mass (flux at 4000 \AA), and a SSP 1.0 Gyr old accounting for 
the remaining 65\% (25\%)\footnote{See \cite{cid05} for the robustness of Starlight 
in recovering different SFHs.}.
The simulated spectrum and the resulting SFH are shown in Figure \ref{fig:mara}.
Even if with different fractions of mass, the two components have been detected, constraining the time since quenching (youngest) and the time since the first
important episode of star formation (oldest). 
{Therefore, we are confident that the SFH found for C1-23152 is not affected by 
the major contribution in light from the youngest population.}

The analysis shows that the build-up of C1-23152 has taken place in about
$\Delta t_{build}$$\simeq$600 Myr as constrained
by the oldest stellar component, i.e. in the range 3.35$<$$z$$<$4.6.
The youngest stellar component constrains the time since quenching at 
$t_{quench}$$\sim$150 Myr, in agreement with the upper limit to the current star 
formation rate, SFR$<$6.5 M$_\odot$ yr$^{-1}$ and with the constraints
derived in \S\ \ref{sec:abs}.
It follows that the stellar mass ($\sim$2$\times$10$^{11}$ M$_\odot$, see below)
has been formed in an interval $\Delta t_{SF}$$\simeq$450-500 Myr, 
corresponding to an average $\overline{SFR}$$>$400 M$_\odot$ yr$^{-1}$.
{ If the SFH was characterized by the two main episodes of star formation
(Fig. \ref{fig:sfh}), 
their SFR would be $\sim$1200 M$_\odot$ yr$^{-1}$}.\footnote{We used eq. (5) in \cite{asari07} and we derived the SFR
as the ratio between the fraction of mass in the burst by the age bin size,
considering 
that the age resolution of the models is 100 Myr for age$>$500 Myr and 50 Myr at younger ages.
Therefore, this SFR is a lower limit driven by the age resolution of the models. }
}

{We also adopted an independent approach by using FAST++ 
\citep{fast18,schreiber18}\footnote{https://github.com/cschreib/fastpp} 
to perform a simultaneous parametric fitting of the 
spectrum and the multi-wavelength UltraVISTA photometry.
The fitting is described in detail in appendix \ref{sec:fast}.
According to the adopted SFHs (a delayed exponentially declining SFH
and a double exponentially increasing/decresing SFH),
the build-up of the galaxy would be realized in 
less than 250 Myr prior to observations (mass assembly 3.35$<$$z$$<$3.8),
a time significantly shorter than the one obtained through FSF.
The resulting mean stellar age is younger than the ages
obtained through ALF and \texttt{STARLIGHT}.
This approach deals principally with broad/narrow-band photometry, 
i.e. with the stars producing the dominant light.
Therefore, in this case, the youngest stellar population could 
affect the results \citep[see e.g.,][]{maraston10,greggio11}.
}

\subsection{Stellar mass estimate}
To estimate the stellar mass of C1-23152, we adopted different approaches: 
(i) we relied on the \texttt{STARLIGHT} FSF results, using the relation 
M$^*$=m$_{ini}$$\times$4$\pi$d$^2_L$/M$_\odot$ \citep{cid07},
where m$_{ini}$ is the normalization factor of the EMILES composite model to
obtain the observed flux and d$_L$ is the luminosity distance.
This approach provides M$^*$=1.5($\pm0.2$)$\times$10$^{11}$ M$_\odot$
after having normalized the spectrum to the observed K-band flux
(K$_{AB}$=20.31) of C1-23152;
(ii) we used the result of ALF (see above).
We normalized the EMILES SSP model corresponding to age$_L$=0.24 Gyr and Z$_L$=0.27
to match the V-band restframe flux of C1-23152 from \cite{marsan15}.
This provides a stellar mass estimate of M$^*$$=$1.0$\times$10$^{11}$ M$_\odot$;
(iii) we used the outputs of FAST++ (see Appendix \ref{sec:fast}) based on BC03 models, which provides values in the range 
M$^*$=2.6-3.1$^{+0.5}_{-0.0}$$\times$10$^{11}$ M$_\odot$
(consistent with the estimate by \cite{marsan15}, M$^*$=2.8$^{+0.6}_{-0.7}$$\times$10$^{11}$, 
based on SED fitting, scaled to Chabrier IMF).
Combining all these values,
we obtain a final estimate of 
M$^*$=2.0($\pm$0.7)$\times$10$^{11}$ M$_\odot$\footnote{For a Salpeter
IMF, given the ages as measured in Tab. \ref{tab:fit} , the stellar mass is a factor 
$\sim$1.3 larger than for a Chabrier IMF (as derived from EMILES models), 
i.e. M$^*$=2.6($\pm$0.7)$\times$10$^{11}$ M$_\odot$.}, consistent with the constraint imposed
by the dynamical mass (see below).
The uncertainty accounts for the different estimates of M$^*$,
based on different methods, software and models.
The resulting stellar mass density within R$_e$ is
$\Sigma_e^{M^*}$=$\Sigma_{1kpc}=3.2(\pm0.7)\times10^{10}$ M$_\odot$ kpc$^{-2}$
where $\Sigma_e^{M^*}$=0.5$M^*$/($\pi$R$^{2}_e$), i.e. we assumed that the stellar mass profile 
follows the luminosity profile of the galaxy.

\begin{table*}
 \caption{\label{tab:sum}Physical properties of C1-23152}
 \centerline{
\begin{tabular}{lcr}
\hline
Parameter& Value & Comment \\
\hline
Redshift, $z$ & 3.352$\pm$0.002 & From  \cite{marsan15}\\
Effective Radius, R$_e$ & 1.0$\pm$0.1 kpc & in F160 band from  \cite{marsan15}\\
Sersic index, $n$ & 4.4 &  in F160 band from \cite{marsan15}\\
Stellar velocity dispersion, $\sigma_e$& 409$\pm$60 km s$^{-1}$ & Scaled to the effective radius\\
Age$_{M^*}$    & 400$^{+30}_{-70}$ Myr & Age of the bulk of stellar mass\\
Metallicity [Z/H]& 0.25$^{+0.006}_{-0.10}$ & Expressed as log(Z/Z$_\odot$)\\
Extinction, A$_V$& 0.05$^{+0.06}_{-0.05}$ mag &  \\
Dynamical mass, $M_{dyn}$ & 2.2$(\pm0.4)\times10^{11}$ M$_\odot$ & From eq. (4) \\
Stellar mass, M$^*$ & 2.0($\pm$0.7)$\times$10$^{11}$ M$_\odot$ & Mean of different estimates\\
Current SFR & $<6.5$ M$_\odot$ yr$^{-1}$ & H$_\beta$ emission line limit\\
Build-up timescale, $\Delta t_{build}$& 600$\pm$100 Myr & 3.35$<$$z$$<$4.6 assembly of galaxy\\
Time since quenching, $t_{quench}$ & 150$\pm$50 Myr & Time since last burst\\
Stellar mass formation, $\Delta t_{SF}$&  450$\pm$110 Myr & Interval of star formation\\
Average SFR within $\Delta t_{SF}$ & 440$\pm$110 M$_\odot$ yr$^{-1}$ & SFR required to form M$^*$ in $\Delta t_{SF}$\\
\hline
\end{tabular}
}
\end{table*}

\subsection{Stellar velocity dispersion measurement and dynamical mass estimate}
A velocity dispersion measurement was performed by fitting the observed spectrum 
using 
\texttt{pPXF} \citep{cappellari04,cappellari17}. 
The fitting was performed by masking out the regions with emission lines ([OII], H$\beta$, [OIII]) 
and the region of low atmospheric transmission (18000-19000 \AA) between H and K bands
(see Fig. \ref{fig:spec}).
{ In the fitting, a Legendre polynomial with degree=4 is added to correct the template 
continuum shape during the fit. 
We verified that the result does not depend on the polynomial degree by varying it in 
the range 1-4.}
The stability of the measurement with respect to the wavelength range considered was tested by 
shifting and varying the width of the masked regions by about 200 \AA, and by fitting the H-band
data only. 
In all the cases considered, we obtained values within 5\% of our nominal $\sigma$ estimate. 
The dependence of the velocity dispersion measurements of the library of templates used, 
was tested by repeating the fitting to the observed spectrum with different libraries of SSP models,
EMILES-Padova \citep{vazdekis15}, BC03 and M11. 
These additional sets of templates provided values within 6\% from the reference 
value.\footnote{Velocity dispersion measurement based primarily on Balmer lines, dominated
by high rotational velocity A-type stars, may be subject to systematics if the correct
stellar population is not matched \citep[e.g.][]{belli17}.
We tested for possible systematics by repeating the fit with stellar spectra
from the Indo-U.S. library \citep{valdes04}, first using only F, G and K stars,
than including also A stars.
We obtained, in the two cases, $\sigma_{obs}$=410$\pm$41 km $s^{-1}$ and 
$\sigma_{obs}$=384$\pm$48 km $s^{-1}$
respectively, with the value obtained using stellar population synthesis models 
($\sigma_{obs}$=396 km $s^{-1}$) in between.} 

The galaxy stellar velocity dispersion $\sigma_*$ was derived from the relation
$\sigma^2_*=\sigma_{obs}^2-\sigma_{inst}^2-\sigma_{stack}^2$,
where $\sigma_{obs}=396$ km $s^{-1}$ is the velocity dispersion resulting from the \texttt{pPXF}
spectral fitting, 
$\sigma_{inst}\sim100$ km $s^{-1}$ is the broadening due to the instrumental resolution,
$\sigma_{stack}\sim$10 km $s^{-1}$ is the broadening 
due to the uncertainty in the wavelength calibration of the frames.
The robustness of $\sigma_{obs}$ estimate was tested by repeating the measurement for a set of 100 
simulated spectra (see Appendix \ref{sec:sim}).  
We estimated $\sigma_*=383\pm60$ km $s^{-1}$ within a diameter aperture of $2r$=0.75 arcsec (the slit width).
The resulting velocity dispersion within the effective radius R$_e$=1 kpc \citep{marsan15} is
$\sigma_e=409\pm60$ km $s^{-1}$, 
where we used the relation  $\sigma_*/\sigma_e=(r/r_e)^{-0.065}$ \citep{jorgensen95,cappellari06}\footnote{This correction is derived from observations of local resolved galaxies. However, 
being a power-law, it can be applied also to spatially unresolved galaxies, as it is the case
for C1-23152.}.

The dynamical mass $M_{dyn}$ of the galaxy has been derived from the velocity dispersion $\sigma_e$
and the effective radius R$_e$ through the relation
\begin{equation}
 M_{dyn}=k_n{{\sigma_e^2 R_e}\over G}
\end{equation}
where G is the gravitational constant and $k_n$ is the virial coefficient which takes 
into account the distribution of both luminous and 
dark matter (DM) and the projection effects \citep{bertin02,lanzoni03}.
We used $k_n=8.88-0.831n+0.0241n^2$ \citep{cappellari06}\footnote{ We note that the
relations relevant to dynamical properties of galaxies heavily rely
on local studies. It is uncertain if these relations can be applied as they are also
to galaxies in the early Universe. However, we noticed that, for a theoretical 
value of K$_n$=5,
the dynamical mass and the DM fraction would change by about 13\%.} with S\'ersic index $n=4.4$, as resulting 
from surface brightness fitting \citep{marsan15}.
We thus obtained $M_{dyn}=2.2\pm0.4\times10^{11}$ M$_\odot$.

We note that, in case of Chabrier IMF, the DM fraction is less than 10\%, 
even if with large errors ($f_{DM}(Cha)=(M_{dyn}-M^*)/M_{dyn}$$=0.09\pm0.3$) while, 
for a Salpeter IMF $f_{DM}(Sal)$=-0.18$\pm$0.3, still consistent 
with a null fraction.
These results suggest that the DM fraction within the effective radius is extremely low
in massive dense ETGs, as already noticed in ETGs at $z\sim1.3$ \citep{saracco20},
and that a bottom-heavy IMF (higher ratio of low- to high-mass stars) 
is disfavored in dense ETGs, irrespective of their redshift \citep{gargiulo15}.
Notice that, based on results at $z$$\sim$0 
a bottom-heavy IMF is expected to be confined in the very central regions of galaxies, making the integrated stellar mass-to-light ratio within the effective radius below
the expectation for a Salpeter IMF \citep[see, e.g.][]{labarbera19}. 
However, our observations virtually collect the whole object light and cannot detect
this effect.

\section{Discussion and conclusions}
\begin{figure*}
\centerline{
 \includegraphics[width=19truecm]{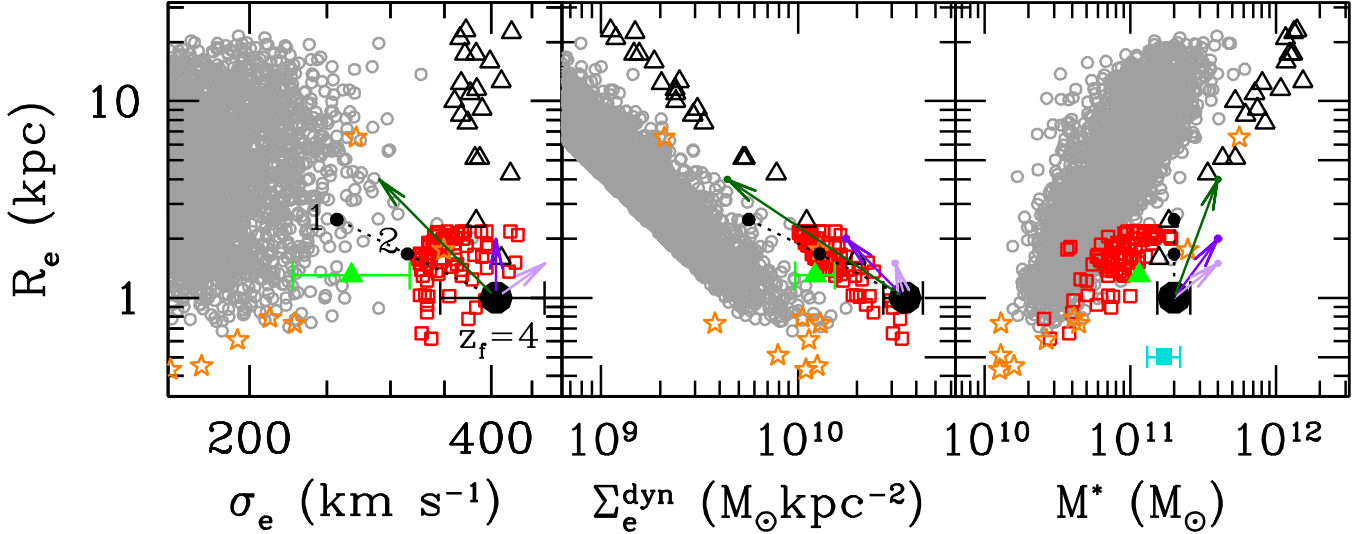}
 }
\caption{\label{fig:z0} Physical properties of C1-23152 and of ETGs in the local Universe.
The half light radius R$_e$ of galaxy C1-23152 (black filled dot) is plotted against:
(from left to right) velocity dispersion $\sigma_e$, surface mass density $\Sigma^{dyn}_e$
and stellar mass M$^*$, and compared to the same parameters of other ETGs in the local Universe
(open symbols) and at redshift $z>3$ (filled symbols). 
For the local Universe, the gray circles represent ETGs selected from SPIDER sample
\protect\citep{labarbera10}, the black triangles are ETGs with $\sigma_e$$>$350 km s$^{-1}$
\protect\citep{bernardi06}, the red squares are compact ETGs with high velocity dispersion \protect\citep{saulder15}
and the orange stars are compact ETGs missed by the Sloan Digital Sky Survey 
\protect\citep[SDSS;][]{damjanov15}.
At redshift $z>3$, the properties of C1-23152 are compared with those of a galaxy 
 at $z\simeq4$ \protect\citep{tanaka19,tanaka20} (green filled triangle), and an ETG
at $z\simeq3.7$ \protect\citep{glazebrook17} (light-blue square).
The arrows indicate the paths for C1-23152 to grow in stellar mass by a factor 2 
from $z$=3.35 to $z$=0 by experiencing: a wet-merger (light-purple arrow) 
during which a significant fraction of the additional stellar mass is formed through dissipative processes; a major equal mass $\sim$(1:1) dry-merger (purple arrow);
accretion of small systems (dark green) with mass ratio $\sim$(1:10) (see text).
The small black dots connected by a dashed line represent the properties that a C1-23152-like
galaxy would have if it forms at redshift $1<z_f<4$, assuming that its properties (equal
to those of C1-23152 at $z$=4) scale according
to the matter density of the Universe, where we assumed $\rho_m(z)\propto(1+z)^3$.
}
\end{figure*}
\subsection{The formation of a massive ETG at high-z}
The analysis of the LBT-LUCI spectrum of the galaxy C1-23152 has lead
to the following results (summarized in Tab.\ref{tab:sum}).
C1-23152 is an early-type galaxy hosting an AGN at $z=3.35$ that assembled  
2.0($\pm$0.5)$\times10^{11}$ M$_\odot$ of stars, quenched its star formation and 
shaped its morphology in the $\sim$600 Myr preceding the observation, i.e. 
between 3.35$<$$z$$<$4.6, as constrained by the oldest stellar component
detected in the galaxy.
This fast build-up is significantly shorter than that expected based on dynamical 
friction merging timescales, larger than 2-3 Gyr \citep[e.g.,][]{boylan08}.
The short assembly time suggests that the stellar mass growth has taken place in-situ 
through highly dissipative processes rather than ex-situ through mergers of pre-existing stellar systems \cite[e.g.,][]{kroupa20}.

The high stellar velocity dispersion, $\sigma_e$=409$\pm$60 km s$^{-1}$, confirms
the high mass, M$_{dyn}$=$2.2(\pm0.4$)$\times$10$^{11}$ M$_\odot$,
and the high mass density of the galaxy.
The surface stellar mass density within R$_e$ is
$\Sigma_e^{M^*}$=$\Sigma_{1kpc}=3.2(\pm0.7$)$\times$10$^{10}$ M$_\odot$ kpc$^{-2}$
($\Sigma_e^{M_{dyn}}$=$3.5(\pm0.5$)$\times$10$^{10}$ M$_\odot$ kpc$^{-2}$), 
comparable with the density of the densest and most 
massive ETGs in the local Universe, as shown in Figure \ref{fig:z0}.
This high mass density coherently argues for a highly dissipative formation process
as expected theoretically \citep[e.g.,][]{hopkins10a,lapi18}, and
from the empirical Kennicutt-Schmidt law \citep{kennicutt98a}.
This fast dissipative process could result from a cosmological merger of 
gas-dominated systems \citep[e.g.][]{naab07} or from violent  instability of a gas-rich disk 
\citep{dekel14}, 
and points toward a role of the density in regulating the quenching process \citep{woo15,tacchella16}.

{ The whole stellar mass observed at $z=3.35$ has been formed within a short interval,
$\Delta t_{SF}\simeq$450 Myr, corresponding to an average 
$\overline{SFR}$$>$400 M$_\odot$ yr$^{-1}$,
 possibly through two main episodes of star formation 
or through a continuous episode}.
The resulting stellar population has a mean mass-weighted 
Age$_M$=400$^{+30}_{-70}$ Myr.
The supersolar metallicity, [Z/H]=0.25$^{+0.006}_{-0.10}$  agrees with a fast
dissipative process and points toward a star formation efficiency much higher 
than the replenishment time of the gas, resembling a nearly closed-box enrichment.
{ Supersolar metallicity ([Z/H]=0.13$^{+0.10}_{-0.08}$)  is also obtained when
 $\alpha$-enhanced models are considered.
 These abundance ratios may better match the overabundance of [Mg/Fe] observed in
 local massive ETGs \citep[e.g.,][]{worthey92}, 
 overabundance usually interpreted as a result of short formation timescales 
 \citep[$<$1 Gyr; see e.g.,][]{trager00,renzini06}.}
Indeed, \cite{vazdekis96} show that, in case of closed-box regime, supersolar metallicity 
can be reached just within the first 100-200 Myr of star formation, a timescale
comparable to the star formation of C1-23152 \cite[see also][]{pantoni19}.
It is interesting to note that, in this approximation, the expected final extinction 
is negligible, 
as indicated by the virtually null value (A$_V$=0.05$\pm$0.05) obtained by spectral fitting. 

The star formation in C1-23152 has halted $\sim$100-150 Myr before observation,
leaving a residual star formation $<$6.5 M$_\odot$ yr$^{-1}$.
Therefore, quenching mechanism must have been extremely efficient to reduce the star
formation to few M$_\odot$ yr$^{-1}$ in less than 150 Myr. 
The presence of an AGN could explain the fast quenching even if it is 
usually associated to a powerful outflow  \citep[e.g.,][]{cicone14,maiolino12} whose presence is difficult 
to assess from our data.
Our analysis cannot establish a causality between the high mass density or the
AGN and the extremely fast suppression of star formation.
On the other hand, this study suggests that they can play a role in the very fast
quenching process of massive galaxies in the early Universe.

\subsection{The evolution of a massive ETG since z=3.35}
Figure \ref{fig:z0} compares the physical properties (effective radius, velocity dispersion,
surface mass density and stellar mass) of C1-23152 (big black points) with those of 
local ETGs with comparable mass (see \S\ 2).
{ In particular, for the local Universe, we considered ETGs selected 
from SPIDER sample \citep[gray circles;][]{labarbera10}, the ETGs with 
$\sigma_e$$>$350 km s$^{-1}$ from \cite{bernardi06} (black triangles), 
the compact ETGs with high velocity dispersion  selected by \cite{saulder15} (red square) 
and the compact ETGs missed by the Sloan Digital Sky Survey (SDSS) and identified by 
\cite{damjanov15} (orange stars).
At redshift $z>3$, the properties of C1-23152 are compared with those of the galaxy 
 at $z\simeq4$ \citep{tanaka19,tanaka20} (green filled triangle), and an ETG
at $z\simeq3.7$ \citep{glazebrook17} (light-blue square).}

The evolutionary path that C1-23152 will follow in the subsequent $\sim$12 Gyr, whether
minor merger will increase its size, and/or a secondary burst of star formation will
rejuvenate the mean stellar age, or a major merger will increase its mass and size,
or it will arrive unperturbed to $z=0$, cannot be univocally determined.
However, we investigated the properties of the possible descendant of C1-23152 
according to the different evolutionary paths that the galaxy could follow. 

{ In an equal mass (1:1) dry-merger, the growth in stellar mass is approximately the  
same  as  the growth in size \citep[e.g.,][]{ciotti07,hopkins09a}.
Consequently the velocity dispersion remains nearly constant,
and the surface mass density decreases as the size increase.
This case is shown in Figure \ref{fig:z0} by the purple arrow for a mass increase by a factor 2,
i.e. for a single major merger as expected for high-mass galaxies \citep[e.g.][]{delucia06}.

If a mass increase by a factor 2 occurs via major wet-merger during which a 
significant fraction of the additional stellar mass is formed in-situ through 
star formation, then the size of the resulting galaxy 
increases less than a factor 2, depending on the fraction of mass involved in the dissipative 
process \citep{ciotti07,hopkins09a}.
Consequently, the velocity dispersion slightly increases, as the square root of the ratio between 
the mass and the size of the final system, while the surface density scales as the ratio
between mass and the square of the radius. 
This case is shown by the light-purple arrow in Figure \ref{fig:z0}, representing a size
increase by a factor 1.5.

Finally, if the mass increase is due to accretion of very small systems,
i.e. minor mergers, then the size increase is greater than that observed for
dry major merger scenario.
For a mass increase by a factor 2 and assuming a single merger of mass ratio 
$\sim$(1:10), the radius 
grows by a factor 4 while the velocity dispersion and the 
stellar mass surface density decrease by a factor of $\sim$1.4 
and 8, respectively \citep{ciotti07,hopkins09a,naab09,bezanson09}.
This case is shown by the dark-green arrow in Figure \ref{fig:z0}.

The cases considered show that C1-23152 will hardly  
grow its mass significantly through dissipative processes since the resulting 
galaxy would have a mass density and velocity dispersion higher than those
of the densest and highest velocity dispersion ETGs in the local Universe.
On the other hand, a mass growth through dry major or minor merging is instead conceivable
\citep[but see][for the large scatter introduced by dry mergers in the scaling relations]{nipoti09}.

It is worth noting that \cite{marchesini14}, using a semi-empirical approach based on 
abundance matching, find that the likely progenitors of local ultra-massive galaxies 
(log(M$^*$)$\simeq$11.8 M$_\odot$) growth by 0.56$^{+0.35}_{-0.25}$dex in stellar mass
since $z$=3,
i.e., their mass at $z$$\sim$3 is 2-3$\times$10$^{11}$ M$_\odot$, consistent with the mass 
of C1-23152 \citep[see also][for a discussion on the size evolution of galaxies since $z$$\sim$4]{kubo18}.
}

Finally, Figure \ref{fig:z0} shows also that the densest massive ETGs in the local Universe share 
the same extreme physical properties of C1-23152 at $z=3.352$.
Theoretically, it is established that, at all redshifts, dissipation is the most
important factor determining the mass density and, hence, size of galaxies \citep[e.g.,][]{hopkins09a}.
Given the higher density of the Universe and the larger gas fractions at high redshift, 
densest galaxies are expected to form in the early Universe \citep{hopkins10a}.
Accordingly, lower redshift formation implies less dense galaxies.
In Figure \ref{fig:z0}, the dashed line connecting small black dots represents the 
properties that a C1-23152-like galaxy would have, if it forms at redshift $1<z_f<4$, 
assuming that its properties (equal to those of C1-23152 at $z_f$=4, for simplicity) 
scale as the matter density of the Universe, where we assumed the simple scaling relation 
$\rho_m(z)\propto(1+z)^3$.

If the densest galaxies form at high redshift, they should host the oldest stellar populations.
Indeed, observations have established that, at fixed mass, denser galaxies host 
older stars irrespective of their redshift \citep[e.g.][]{saracco09,saracco11,valentinuzzi10} and, consequently,
their formation epoch precedes the one of less dense galaxies \citep[e.g.,][]{saracco20,estrada20}.
Therefore, even if C1-23152 may follow different possible evolutionary paths, 
the densest massive ETGs in the local Universe most likely have a C1-23152-like galaxy 
as progenitor that arrived unperturbed to $z$=0, since their extreme properties 
require physical conditions 
for their formation similar to those for C1-23152, difficult to realize at lower redshift.

\acknowledgments
We thank the anonymous referee for the useful and constructive comments.
This work is based on observations carried out at the Large Binocular Telescope (LBT) under
the observation program 2017B-C2743-3 (P.I. Paolo Saracco).
The LBT is an international collaboration among institutions in Italy, Germany and United States.
LBT Corporation partners are: Istituto Nazionale di Astrofisica (INAF), Italy; 
LBT Beteiligungsgesellschaft, Germany, representing the Max-Planck Society, 
The Leibniz Institute for Astrophysics Potsdam, and Heidelberg University;
The University of Arizona on behalf of the Arizona Board of Regents;  
The Ohio State University, and The Research Corporation, on behalf of The University of Notre Dame, 
University of Minnesota and University of Virginia.
We acknowledge the support from the LBT-Italian Coordination Facility for the execution
of the observations, the data distribution and for support in data reduction.
PS would like to thank R. Paulson and Mayhem project.
DM and MA acknowledge support by grant number NNX16AN49G issued through the NASA 
Astrophysics Data Analysis Program (ADAP). 
Further support was provided by the Faculty Research Fund (FRF) of Tufts University.
FLB acknowledges support from grant AYA2016-77237-C3-1-P from the Spanish Ministry of 
Economy and Competitiveness (MINECO).
GW acknowledges suppport by the National Science Foundation through grant AST-1517863, by HST program
number GO-15924, and by grant number 80NSSC17K0019 issued through the NASA Astrophysics Data Analysis
Program (ADAP). 
Support for program number GO-15294 was provided by NASA through a grant from the Space Telescope Science
Institue, which is operated by the Association of Universities for Research in Astronomy, Incorporated, 
under NASA contract NAS5-26555.

\appendix

\section{\label{sec:fast} FAST++ fitting}
A great advantage of FAST++ is that the whole SED 
can be considered in the fitting, 
simultaneously modeling the photometry and the spectrum.
However, the best fitting model is chosen among a set of model templates whose SFH 
is defined a priori.
Any SFH implies a trade off between stellar age and duration of the
star formation to best match the data.
Therefore, the resulting age, interval within which the star formation can 
have taken place and SFR depend on the adopted SFHs.
Metallicity is not a free parameter in the fitting, as it is fixed with the choice of adopted models. 
We considered, BC03 models with \cite{chabrier03} IMF and \cite{calzetti00} extinction law. 
Two SFHs were considered: a delayed exponentially declining SFH ($\tau$-mod hereafter)
with e-folding time log($\tau$/yr)=[7-10] with step 0.1, and log(age/yr)=[7-9] with step 0.05, 
and a double exponential \citep[2$\tau$-mod hereafter; e.g.,][]{schreiber18}  
exponentially increasing (log(trise/yr)=[7-9.5] with step 0.5), and exponentially decreasing
(log(tdecl/yr)=[7-9.5] with step 0.5). 
In both cases, a recent burst was allowed to be present. 
For each of these two SFHs, a solar metallicity (Z=0.02) and super-solar metallicity (Z=0.05) were adopted. 
For C1-23152, we fit together the LBT spectrum and the multiwavelength UltraVISTA 
photometry composed of 49 bands.
In addition to the 30 photometric bands used in the construction of the UltraVISTA DR1 catalog of \cite{muzzin13}, photometry from the following images were included: 
ultra-deep optical imaging from HSC \citep{tanaka17}, the UltraVISTA DR3 YJHK 
($\sim$1.2 mag deeper than the DR1 images), deeper (by $\sim$1 mag) IRAC photometry from SPLASH \citep{mehta18} and SMUVS \citep{caputi17, ashby18}, UltraVISTA NB118, 
five CFHTLS deep optical images, and medium-band NIR images from the NMBS 
(see Muzzin et al. 2020, in prep.).

In all the cases, the SED+spectrum excluded the presence of a recent burst.
For Z=Z$_\odot$(2.5Z$_\odot$), in the $\tau$-mod case, the star formation starts 
251(224) Myr prior to observation and 50\% of the mass (2.9(2.6)$\times$10$^{11}$ M$_\odot$)
is formed in about 42(42) Myr at a mean  SFR$\sim$3430(3060) M$_\odot$ yr$^{-1}$. 
The star formation drops to 10\% 123(96) Myr prior observation, leaving a residual
SFR=8(19) M$_\odot$ yr$^{-1}$.
The extinction is A$_V$=0.5(0.4) mag.

The 2$\tau$-mod SFH implies a slightly faster buildup and quenching 
(and larger peak of star formation): for Z=Z$_\odot$(2.5Z$_\odot$),
the star formation starts 224(178) Myr prior to 
observation and 50\% of the mass (3.1(2.7)$\times$10$^{11}$ M$_\odot$) is formed in 38(8) Myr at a mean SFR$\sim$4066(16800)
M$_\odot$ yr$^{-1}$.
The star formation drops to 10\% 135(151) Myr prior to observation, leaving a
residual SFR=25(23) M$_\odot$ yr$^{-1}$.
The extinction is A$_V$=0.6(0.4) mag.

Therefore, according to the adopted SFHs, the build-up of C1-23152 would be realized
in less than 250 Myr, i.e. between 3.35$<$$z$$<$3.8. 
The resulting mean stellar age is younger than the ages obtained 
with \texttt{STARLIGHT} for the same BC03 models (see Tab.\ref{tab:fitmod}),
and with ALF method, and the build-up of the galaxy significantly shorter.

\section{\label{sec:sim} Simulations}
The robustness of age, metallicity and velocity dispersion measurements was assessed by repeating 
the fitting to two sets of 100 simulated spectra having the same S/N of the observed spectrum.
The first set (MC-set hereafter) was obtained through a Monte Carlo approach, i.e. by varying the value 
of the true observed spectrum $f_{obs}(\lambda)$
by a shift $\delta f$ randomly chosen from a gaussian distribution with sigma d$f(\lambda)$, 
the error on $f_{obs}(\lambda)$.
The second set of simulated spectra (R-set hereafter) was obtained 
by summing to the best-fitting model template $f_{mod}(\lambda)$ the residuals 
R($\lambda$)=$f_{obs}(\lambda)$-$f_{mod}(\lambda)$ randomly shuffled in wavelength within windows
200 \AA\ wide, along the wavelength axis, i.e. 
$f_{sim}(\lambda)$=$f_{mod}(\lambda)$+R($\lambda_{RND}$)

The distributions of the best fitting values obtained with the two sets of simulated spectra 
do not differ significantly.
However, R-set has the advantage to take into account possible systematics in 
the observed spectrum due to sky residuals or sky absorptions.
Reshuffling allows for considering the effects of these systematics both in 
the absorption lines fitting and in the full spectral fitting.
For these reasons, we consider the results obtained from R-set as reference. 
\begin{figure*}
 \centerline{
 \includegraphics[width=14truecm]{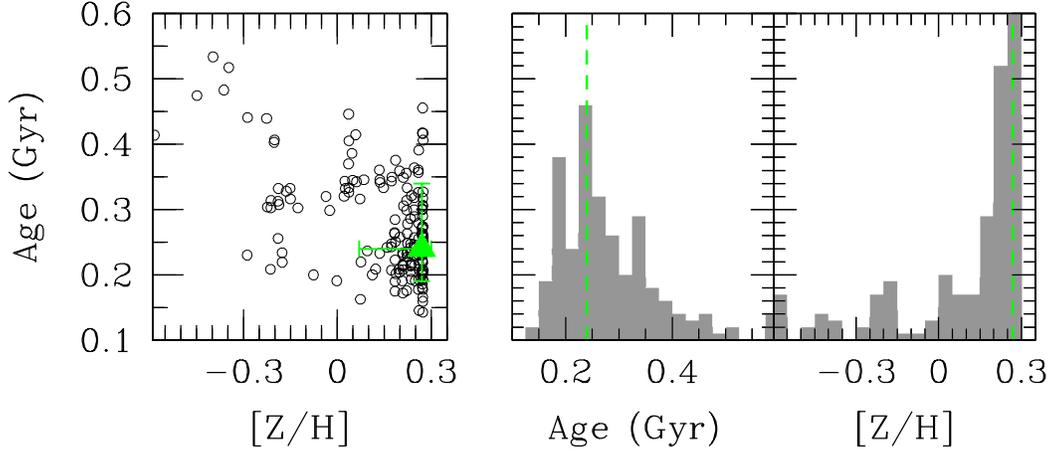}}
\caption{\label{fig:simlab} Results of absorption lines fitting to the 100 simulated spectra.
For each simulated spectrum (see \S\ \ref{sec:sim}), the best fitting age and 
metallicity have been found by comparing the measured line-strengths to those predicted by SSP 
with varying age and metallicity.
The histograms show the distributions of the best-fitting values.
Green filled triangle and dashed lines mark the best-fitting values
obtained for C1-23152 reported in Tab.\ref{tab:fit}.
}
\end{figure*}

\begin{figure*}
\centerline{
 \includegraphics[width=14truecm]{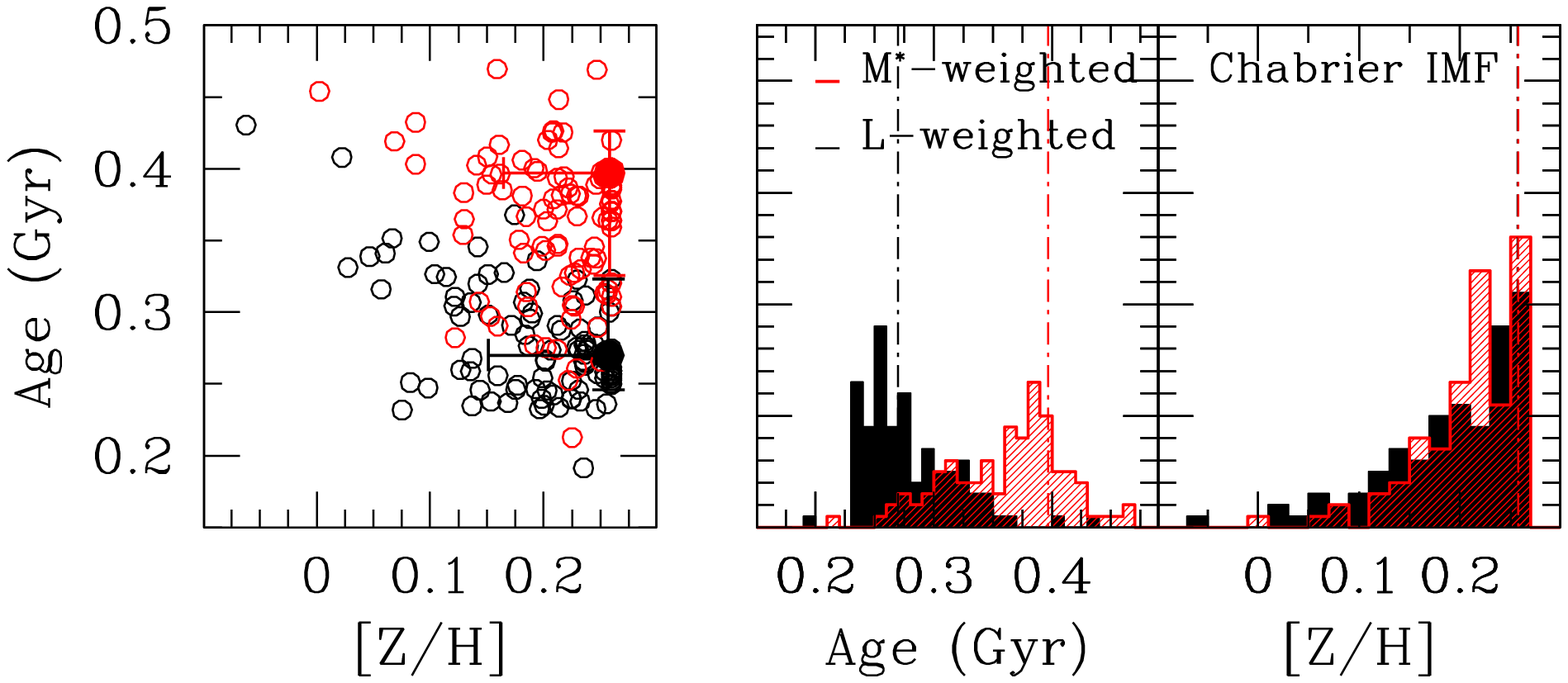}}
 \centerline{
 \includegraphics[width=14truecm]{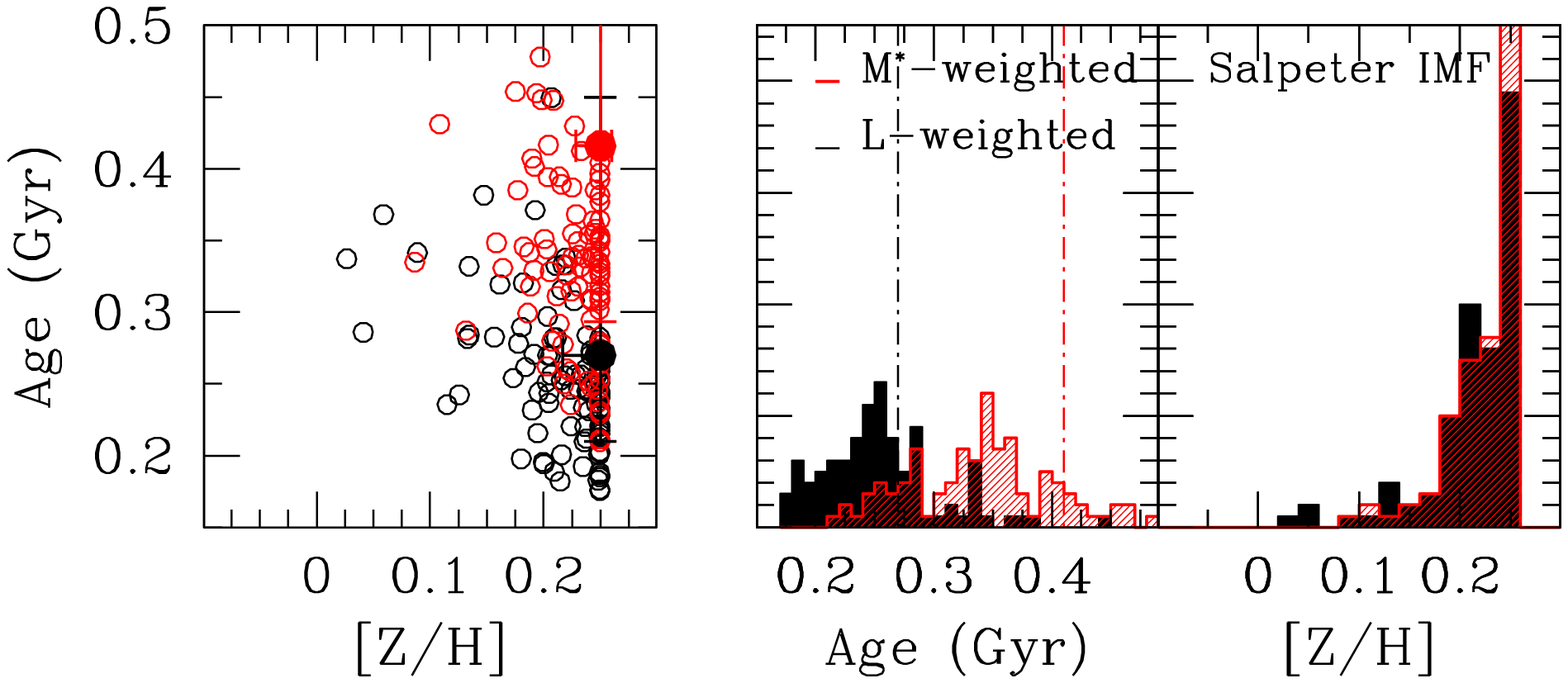}
}
\caption{\label{fig:simstar}Results of full spectral fitting to simulated spectra. 
For each simulated spectrum, we derived age and metallicity (open dots) 
by running \texttt{STARLIGHT} as for the real spectrum. 
The histograms show the distributions of the estimated values.
Black color identifies the light-weighted values, while red color the mass-weighted values. 
Filled dots and dashed lines are the best fitting values obtained 
for galaxy C1-23152 reported in Tab. \ref{tab:fit}. 
The upper panel are the results obtained with a set of EMILES models based on Chabrier IMF; 
the lower panel with a set of models based on Salpeter IMF.
}
\end{figure*}

The results of the absorption lines fitting obtained with R-set 
of simulated spectra is shown in Figure \ref{fig:simlab} while
Figure \ref{fig:simstar} shows the results obtained with full spectral fitting (\texttt{STARLIGHT}). 
To quantify the dependence of the results on the IMF assumed,
for each simulated spectrum we measured age and metallicity by running \texttt{STARLIGHT} 
using the same parameters setup and two bases of SSP models, one based on Chabrier IMF and 
the other on Salpeter IMF, as for the observed spectrum of galaxy C1-23152.
The distribution of the best-fitting extinction values A$_V$ are always constrained within low values, 
with a median extinction A$_V$=0.06$\pm$0.06 mag in the case of Charbier IMF and A$_V$=0.08$\pm$0.09 mag in the case of Salpeter IMF.
Figure \ref{fig:simstar} the results obtained for the two different IMFs.
Salpeter IMF provides best-fitting mass-weighted ages with a slightly larger spread 
with respect to Chabrier IMF.
Notice that the differences due to varying the IMF are not significant.
The errors at 68\%\ confidence level reported in Tab. 1 for galaxy C1-23152 are
derived from the distribution of the best fitting values obtained from these simulations.\\

\bibliographystyle{aasjournal}
\bibliography{paper_vega_Apj_r1}

\begin{thebibliography}{}
\expandafter\ifx\csname natexlab\endcsname\relax\def\natexlab#1{#1}\fi
\providecommand{\url}[1]{\href{#1}{#1}}
\providecommand{\dodoi}[1]{doi:~\href{http://doi.org/#1}{\nolinkurl{#1}}}
\providecommand{\doeprint}[1]{\href{http://ascl.net/#1}{\nolinkurl{http://ascl.net/#1}}}
\providecommand{\doarXiv}[1]{\href{https://arxiv.org/abs/#1}{\nolinkurl{https://arxiv.org/abs/#1}}}

\bibitem[{{Ageorges} {et~al.}(2010){Ageorges}, {Seifert}, {J{\"u}tte},
  {Knierim}, {Lehmitz}, {Germeroth}, {Buschkamp}, {Polsterer}, {Pasquali},
  {Naranjo}, {Gemperlein}, {Hill}, {Feiz}, {Hofmann}, {Laun}, {Lederer},
  {Lenzen}, {Mall}, {Mand el}, {M{\"u}ller}, {Quirrenbach}, {Sch{\"a}ffner},
  {Storz}, \& {Weiser}}]{ageorges10}
{Ageorges}, N., {Seifert}, W., {J{\"u}tte}, M., {et~al.} 2010, Society of
  Photo-Optical Instrumentation Engineers (SPIE) Conference Series, Vol. 7735,
  {LUCIFER1 commissioning at the LBT}, 77351L

\bibitem[{{Asari} {et~al.}(2007){Asari}, {Cid Fernandes}, {Stasi{\'n}ska},
  {Torres-Papaqui}, {Mateus}, {Sodr{\'e}}, {Schoenell}, \& {Gomes}}]{asari07}
{Asari}, N.~V., {Cid Fernandes}, R., {Stasi{\'n}ska}, G., {et~al.} 2007, MNRAS,
  381, 263.
\newblock \doarXiv{0707.3578}

\bibitem[{{Ashby} {et~al.}(2018){Ashby}, {Caputi}, {Cowley}, {Deshmukh},
  {Dunlop}, {Milvang-Jensen}, {Fynbo}, {Muzzin}, {McCracken}, {Le F{\`e}vre},
  {Huang}, \& {Zhang}}]{ashby18}
{Ashby}, M.~L.~N., {Caputi}, K.~I., {Cowley}, W., {et~al.} 2018, \apjs, 237,
  39.
\newblock \doarXiv{1801.02660}

\bibitem[{{Baldwin} {et~al.}(1981){Baldwin}, {Phillips}, \&
  {Terlevich}}]{baldwin81}
{Baldwin}, J.~A., {Phillips}, M.~M., \& {Terlevich}, R. 1981, PASP, 93, 5

\bibitem[{{Balogh} {et~al.}(1999){Balogh}, {Morris}, {Yee}, {Carlberg}, \&
  {Ellingson}}]{balogh99}
{Balogh}, M.~L., {Morris}, S.~L., {Yee}, H.~K.~C., {Carlberg}, R.~G., \&
  {Ellingson}, E. 1999, ApJ, 527, 54

\bibitem[{{Beifiori} {et~al.}(2011){Beifiori}, {Maraston}, {Thomas}, \&
  {Johansson}}]{beifiori11}
{Beifiori}, A., {Maraston}, C., {Thomas}, D., \& {Johansson}, J. 2011, A\&A,
  531, A109.
\newblock \doarXiv{1012.3428}

\bibitem[{{Belli} {et~al.}(2014){Belli}, {Newman}, \& {Ellis}}]{belli14}
{Belli}, S., {Newman}, A.~B., \& {Ellis}, R.~S. 2014, ApJ, 783, 117.
\newblock \doarXiv{1311.3317}

\bibitem[{{Belli} {et~al.}(2017){Belli}, {Genzel}, {F{\"o}rster Schreiber},
  {Wisnioski}, {Wilman}, {Wuyts}, {Mendel}, {Beifiori}, {Bender}, {Brammer},
  {Burkert}, {Chan}, {Davies}, {Davies}, {Fabricius}, {Fossati}, {Galametz},
  {Lang}, {Lutz}, {Momcheva}, {Nelson}, {Saglia}, {Tacconi}, {Tadaki},
  {{\"U}bler}, \& {van Dokkum}}]{belli17}
{Belli}, S., {Genzel}, R., {F{\"o}rster Schreiber}, N.~M., {et~al.} 2017, ApJl,
  841, L6.
\newblock \doarXiv{1703.07778}

\bibitem[{{Bernardi} {et~al.}(2006){Bernardi}, {Nichol}, {Sheth}, {Miller}, \&
  {Brinkmann}}]{bernardi06}
{Bernardi}, M., {Nichol}, R.~C., {Sheth}, R.~K., {Miller}, C.~J., \&
  {Brinkmann}, J. 2006, AJ, 131, 1288

\bibitem[{{Bertin} {et~al.}(2002){Bertin}, {Ciotti}, \& {Del
  Principe}}]{bertin02}
{Bertin}, G., {Ciotti}, L., \& {Del Principe}, M. 2002, A\&A, 386, 149

\bibitem[{{Bezanson} {et~al.}(2009){Bezanson}, {van Dokkum}, {Tal},
  {Marchesini}, {Kriek}, {Franx}, \& {Coppi}}]{bezanson09}
{Bezanson}, R., {van Dokkum}, P.~G., {Tal}, T., {et~al.} 2009, ApJ, 697, 1290.
\newblock \doarXiv{0903.2044}

\bibitem[{{Boylan-Kolchin} {et~al.}(2008){Boylan-Kolchin}, {Ma}, \&
  {Quataert}}]{boylan08}
{Boylan-Kolchin}, M., {Ma}, C.-P., \& {Quataert}, E. 2008, MNRAS, 383, 93.
\newblock \doarXiv{0707.2960}

\bibitem[{{Brooks} \& {Christensen}(2016)}]{brooks16}
{Brooks}, A., \& {Christensen}, C. 2016, in Astrophysics and Space Science
  Library, Vol. 418, Galactic Bulges, ed. E.~{Laurikainen}, R.~{Peletier}, \&
  D.~{Gadotti}, 317.
\newblock \doarXiv{1511.04095}

\bibitem[{{Bruzual}(1983)}]{bruzual83}
{Bruzual}, A.~G. 1983, ApJS, 53, 497

\bibitem[{{Bruzual} \& {Charlot}(2003)}]{bruzual03}
{Bruzual}, G., \& {Charlot}, S. 2003, MNRAS, 344, 1000

\bibitem[{{Calzetti} {et~al.}(2000){Calzetti}, {Armus}, {Bohlin}, {Kinney},
  {Koornneef}, \& {Storchi-Bergmann}}]{calzetti00}
{Calzetti}, D., {Armus}, L., {Bohlin}, R.~C., {et~al.} 2000, ApJ, 533, 682

\bibitem[{{Cappellari}(2017)}]{cappellari17}
{Cappellari}, M. 2017, MNRAS, 466, 798.
\newblock \doarXiv{1607.08538}

\bibitem[{{Cappellari} \& {Emsellem}(2004)}]{cappellari04}
{Cappellari}, M., \& {Emsellem}, E. 2004, PASP, 116, 138

\bibitem[{{Cappellari} {et~al.}(2006){Cappellari}, {Bacon}, {Bureau}, {Damen},
  {Davies}, {de Zeeuw}, {Emsellem}, {Falc{\'o}n-Barroso}, {Krajnovi{\'c}},
  {Kuntschner}, {McDermid}, {Peletier}, {Sarzi}, {van den Bosch}, \& {van de
  Ven}}]{cappellari06}
{Cappellari}, M., {Bacon}, R., {Bureau}, M., {et~al.} 2006, MNRAS, 366, 1126

\bibitem[{{Caputi} {et~al.}(2017){Caputi}, {Deshmukh}, {Ashby}, {Cowley},
  {Bisigello}, {Fazio}, {Fynbo}, {Le F{\`e}vre}, {Milvang-Jensen}, \&
  {Ilbert}}]{caputi17}
{Caputi}, K.~I., {Deshmukh}, S., {Ashby}, M.~L.~N., {et~al.} 2017, \apj, 849,
  45.
\newblock \doarXiv{1705.06179}

\bibitem[{{Cardelli} {et~al.}(1989){Cardelli}, {Clayton}, \&
  {Mathis}}]{cardelli89}
{Cardelli}, J.~A., {Clayton}, G.~C., \& {Mathis}, J.~S. 1989, ApJ, 345, 245

\bibitem[{{Cassata} {et~al.}(2013){Cassata}, {Giavalisco}, {Williams}, {Guo},
  {Lee}, {Renzini}, {Ferguson}, {Faber}, {Barro}, {McIntosh}, \&
  coauthors}]{cassata13}
{Cassata}, P., {Giavalisco}, M., {Williams}, C.~C., {et~al.} 2013, ApJ, 775,
  106.
\newblock \doarXiv{1303.2689}

\bibitem[{{Chabrier}(2003)}]{chabrier03}
{Chabrier}, G. 2003, PASP, 115, 763

\bibitem[{{Cicone} {et~al.}(2014){Cicone}, {Maiolino}, {Sturm},
  {Graci{\'a}-Carpio}, {Feruglio}, {Neri}, {Aalto}, {Davies}, {Fiore},
  {Fischer}, {Garc{\'\i}a-Burillo}, {Gonz{\'a}lez-Alfonso}, {Hailey-Dunsheath},
  {Piconcelli}, \& {Veilleux}}]{cicone14}
{Cicone}, C., {Maiolino}, R., {Sturm}, E., {et~al.} 2014, A\&A, 562, A21.
\newblock \doarXiv{1311.2595}

\bibitem[{{Cid Fernandes} {et~al.}(2007){Cid Fernandes}, {Asari}, {Sodr{\'e}},
  {Stasi{\'n}ska}, {Mateus}, {Torres-Papaqui}, \& {Schoenell}}]{cid07}
{Cid Fernandes}, R., {Asari}, N.~V., {Sodr{\'e}}, L., {et~al.} 2007, MNRAS,
  375, L16

\bibitem[{{Cid Fernandes} {et~al.}(2005){Cid Fernandes}, {Mateus}, {Sodr{\'e}},
  {Stasi{\'n}ska}, \& {Gomes}}]{cid05}
{Cid Fernandes}, R., {Mateus}, A., {Sodr{\'e}}, L., {Stasi{\'n}ska}, G., \&
  {Gomes}, J.~M. 2005, MNRAS, 358, 363

\bibitem[{{Cimatti} {et~al.}(2008){Cimatti}, {Cassata}, {Pozzetti}, {Kurk},
  {Mignoli}, {Renzini}, {Daddi}, {Bolzonella}, {Brusa}, {Rodighiero},
  {Dickinson}, {Franceschini}, {Zamorani}, {Berta}, {Rosati}, \&
  {Halliday}}]{cimatti08}
{Cimatti}, A., {Cassata}, P., {Pozzetti}, L., {et~al.} 2008, A$\&$A, 482, 21.
\newblock \doarXiv{0801.1184}

\bibitem[{{Ciotti} {et~al.}(2007){Ciotti}, {Lanzoni}, \&
  {Volonteri}}]{ciotti07}
{Ciotti}, L., {Lanzoni}, B., \& {Volonteri}, M. 2007, ApJ, 658, 65

\bibitem[{{Damjanov} {et~al.}(2015){Damjanov}, {Geller}, {Zahid}, \&
  {Hwang}}]{damjanov15}
{Damjanov}, I., {Geller}, M.~J., {Zahid}, H.~J., \& {Hwang}, H.~S. 2015, ApJ,
  806, 158.
\newblock \doarXiv{1501.04976}

\bibitem[{{Davidge} \& {Clark}(1994)}]{davidge94}
{Davidge}, T.~J., \& {Clark}, C.~C. 1994, \aj, 107, 946

\bibitem[{{Davies}(2007)}]{davies07}
{Davies}, R.~I. 2007, MNRAS, 375, 1099.
\newblock \doarXiv{astro-ph/0612257}

\bibitem[{{De Lucia} {et~al.}(2006){De Lucia}, {Springel}, {White}, {Croton},
  \& {Kauffmann}}]{delucia06}
{De Lucia}, G., {Springel}, V., {White}, S.~D.~M., {Croton}, D., \&
  {Kauffmann}, G. 2006, MNRAS, 366, 499

\bibitem[{{Dekel} \& {Burkert}(2014)}]{dekel14}
{Dekel}, A., \& {Burkert}, A. 2014, MNRAS, 438, 1870.
\newblock \doarXiv{1310.1074}

\bibitem[{{D'Eugenio} {et~al.}(2020){D'Eugenio}, {Daddi}, {Gobat},
  {Strazzullo}, {Lustig}, {Delvecchio}, {Jin}, {Puglisi}, {Calabr{\'o}},
  {Mancini}, {Dickinson}, {Cimatti}, \& {Onodera}}]{deugenio20}
{D'Eugenio}, C., {Daddi}, E., {Gobat}, R., {et~al.} 2020, \apjl, 892, L2.
\newblock \doarXiv{2003.04342}

\bibitem[{{Estrada-Carpenter} {et~al.}(2020){Estrada-Carpenter}, {Papovich},
  {Momcheva}, {Brammer}, {Simons}, {Bridge}, {Cleri}, {Ferguson},
  {Finkelstein}, {Giavalisco}, {Jung}, {Matharu}, {Trump}, \&
  {Weiner}}]{estrada20}
{Estrada-Carpenter}, V., {Papovich}, C., {Momcheva}, I., {et~al.} 2020, arXiv
  e-prints, arXiv:2005.12289.
\newblock \doarXiv{2005.12289}

\bibitem[{{Forrest} {et~al.}(2020{\natexlab{a}}){Forrest}, {Annunziatella},
  {Wilson}, {Marchesini}, {Muzzin}, {Cooper}, {Marsan}, {McConachie}, {Chan},
  {Gomez}, {Kado-Fong}, {Barbera}, {Labb{\'e}}, {Lange-Vagle}, {Nantais},
  {Nonino}, {Pe{\~n}a}, {Saracco}, {Stefanon}, \& {van der Burg}}]{forrest20a}
{Forrest}, B., {Annunziatella}, M., {Wilson}, G., {et~al.} 2020{\natexlab{a}},
  \apjl, 890, L1.
\newblock \doarXiv{1910.10158}

\bibitem[{{Forrest} {et~al.}(2020{\natexlab{b}}){Forrest}, {Marsan},
  {Annunziatella}, {Wilson}, {Muzzin}, {Marchesini}, {Cooper}, {Chan},
  {McConachie}, {Gomez}, {Kado-Fong}, {Barbera}, {Lange-Vagle}, {Nantais},
  {Nonino}, {Saracco}, {Stefanon}, \& {van der Burg}}]{forrest20b}
{Forrest}, B., {Marsan}, Z.~C., {Annunziatella}, M., {et~al.}
  2020{\natexlab{b}}, \apj, 903, 47.
\newblock \doarXiv{2009.07281}

\bibitem[{{Francis} {et~al.}(1991){Francis}, {Hewett}, {Foltz}, {Chaffee},
  {Weymann}, \& {Morris}}]{francis91}
{Francis}, P.~J., {Hewett}, P.~C., {Foltz}, C.~B., {et~al.} 1991, ApJ, 373, 465

\bibitem[{{Gallazzi} {et~al.}(2006){Gallazzi}, {Charlot}, {Brinchmann}, \&
  {White}}]{gallazzi06}
{Gallazzi}, A., {Charlot}, S., {Brinchmann}, J., \& {White}, S. D.~M. 2006,
  MNRAS, 370, 1106.
\newblock \doarXiv{astro-ph/0605300}

\bibitem[{{Gargiulo} {et~al.}(2015){Gargiulo}, {Saracco}, {Longhetti},
  {Tamburri}, {Lonoce}, \& {Ciocca}}]{gargiulo15}
{Gargiulo}, A., {Saracco}, P., {Longhetti}, M., {et~al.} 2015, A\&A, 573, A110.
\newblock \doarXiv{1410.5229}

\bibitem[{{Glazebrook} {et~al.}(2017){Glazebrook}, {Schreiber}, {Labb{\'e}},
  {Nanayakkara}, {Kacprzak}, {Oesch}, {Papovich}, {Spitler}, {Straatman},
  {Tran}, \& {Yuan}}]{glazebrook17}
{Glazebrook}, K., {Schreiber}, C., {Labb{\'e}}, I., {et~al.} 2017, Nature, 544,
  71.
\newblock \doarXiv{1702.01751}

\bibitem[{{Gobat} {et~al.}(2012){Gobat}, {Strazzullo}, {Daddi}, {Onodera},
  {Renzini}, {B{\'e}thermin}, {Dickinson}, {Carollo}, \& {Cimatti}}]{gobat12}
{Gobat}, R., {Strazzullo}, V., {Daddi}, E., {et~al.} 2012, \apjl, 759, L44.
\newblock \doarXiv{1210.4299}

\bibitem[{{Goto} {et~al.}(2003){Goto}, {Nichol}, {Okamura}, {Sekiguchi},
  {Miller}, {Bernardi}, {Hopkins}, {Tremonti}, {Connolly}, {Castander},
  {Brinkmann}, {Fukugita}, {Harvanek}, {Ivezic}, {Kleinman}, {Krzesinski},
  {Long}, {Loveday}, {Neilsen}, {Newman}, {Nitta}, {Snedden}, \&
  {Subbarao}}]{goto03a}
{Goto}, T., {Nichol}, R.~C., {Okamura}, S., {et~al.} 2003, \pasj, 55, 771.
\newblock \doarXiv{astro-ph/0301305}

\bibitem[{{Greggio} \& {Renzini}(2011)}]{greggio11}
{Greggio}, L., \& {Renzini}, A. 2011, {Stellar Populations. A User Guide from
  Low to High Redshift}

\bibitem[{{Hill} {et~al.}(2017){Hill}, {Muzzin}, {Franx}, {Clauwens},
  {Schreiber}, {Marchesini}, {Stefanon}, {Labbe}, {Brammer}, {Caputi}, {Fynbo},
  {Milvang-Jensen}, {Skelton}, {van Dokkum}, \& {Whitaker}}]{hill17}
{Hill}, A.~R., {Muzzin}, A., {Franx}, M., {et~al.} 2017, \apj, 837, 147.
\newblock \doarXiv{1702.06126}

\bibitem[{{Hopkins} {et~al.}(2009){Hopkins}, {Hernquist}, {Cox}, {Keres}, \&
  {Wuyts}}]{hopkins09a}
{Hopkins}, P.~F., {Hernquist}, L., {Cox}, T.~J., {Keres}, D., \& {Wuyts}, S.
  2009, \apj, 691, 1424, \dodoi{10.1088/0004-637X/691/2/1424}

\bibitem[{{Hopkins} {et~al.}(2010){Hopkins}, {Murray}, {Quataert}, \&
  {Thompson}}]{hopkins10a}
{Hopkins}, P.~F., {Murray}, N., {Quataert}, E., \& {Thompson}, T.~A. 2010,
  \mnras, 401, L19.
\newblock \doarXiv{0908.4088}

\bibitem[{{Je{\v{r}}{\'a}bkov{\'a}} {et~al.}(2018){Je{\v{r}}{\'a}bkov{\'a}},
  {Hasani Zonoozi}, {Kroupa}, {Beccari}, {Yan}, {Vazdekis}, \&
  {Zhang}}]{jerabkova18}
{Je{\v{r}}{\'a}bkov{\'a}}, T., {Hasani Zonoozi}, A., {Kroupa}, P., {et~al.}
  2018, \aap, 620, A39.
\newblock \doarXiv{1809.04603}

\bibitem[{{J{\o}rgensen} {et~al.}(1995){J{\o}rgensen}, {Franx}, \&
  {Kjaergaard}}]{jorgensen95}
{J{\o}rgensen}, I., {Franx}, M., \& {Kjaergaard}, P. 1995, MNRAS, 276, 1341

\bibitem[{{Kennicutt}(1998{\natexlab{a}})}]{kennicutt98a}
{Kennicutt}, Robert~C., J. 1998{\natexlab{a}}, ApJ, 498, 541.
\newblock \doarXiv{astro-ph/9712213}

\bibitem[{{Kennicutt}(1998{\natexlab{b}})}]{kennicutt98}
{Kennicutt}, Jr., R.~C. 1998{\natexlab{b}}, ARA\&A, 36, 189

\bibitem[{{Kriek} {et~al.}(2009){Kriek}, {van Dokkum}, {Labb{\'e}}, {Franx},
  {Illingworth}, {Marchesini}, \& {Quadri}}]{kriek09}
{Kriek}, M., {van Dokkum}, P.~G., {Labb{\'e}}, I., {et~al.} 2009, ApJ, 700,
  221.
\newblock \doarXiv{0905.1692}

\bibitem[{{Kriek} {et~al.}(2018){Kriek}, {van Dokkum}, {Labb{\'e}}, {Franx},
  {Illingworth}, {Marchesini}, {Quadri}, {Aird}, {Coil}, \&
  {Georgakakis}}]{fast18}
---. 2018, {FAST: Fitting and Assessment of Synthetic Templates}.
\newblock \doeprint{1803.008}

\bibitem[{{Kroupa} {et~al.}(2020){Kroupa}, {Subr}, {Jerabkova}, \&
  {Wang}}]{kroupa20}
{Kroupa}, P., {Subr}, L., {Jerabkova}, T., \& {Wang}, L. 2020, \mnras, 498,
  5652.
\newblock \doarXiv{2007.14402}

\bibitem[{{Kubo} {et~al.}(2018){Kubo}, {Tanaka}, {Yabe}, {Toft}, {Stockmann},
  \& {G{\'o}mez-Guijarro}}]{kubo18}
{Kubo}, M., {Tanaka}, M., {Yabe}, K., {et~al.} 2018, \apj, 867, 1.
\newblock \doarXiv{1810.00543}

\bibitem[{{La Barbera} {et~al.}(2010){La Barbera}, {de Carvalho}, {de La Rosa},
  {Lopes}, {Kohl-Moreira}, \& {Capelato}}]{labarbera10}
{La Barbera}, F., {de Carvalho}, R.~R., {de La Rosa}, I.~G., {et~al.} 2010,
  MNRAS, 408, 1313

\bibitem[{{La Barbera} {et~al.}(2013){La Barbera}, {Ferreras}, {Vazdekis}, {de
  la Rosa}, {de Carvalho}, {Trevisan}, {Falc{\'o}n-Barroso}, \&
  {Ricciardelli}}]{labarbera13}
{La Barbera}, F., {Ferreras}, I., {Vazdekis}, A., {et~al.} 2013, MNRAS, 433,
  3017.
\newblock \doarXiv{1305.2273}

\bibitem[{{La Barbera} {et~al.}(2019){La Barbera}, {Vazdekis}, {Ferreras},
  {Pasquali}, {Allende Prieto}, {Mart{\'\i}n-Navarro}, {Aguado}, {de Carvalho},
  {Rembold}, {Falc{\'o}n-Barroso}, \& {van de Ven}}]{labarbera19}
{La Barbera}, F., {Vazdekis}, A., {Ferreras}, I., {et~al.} 2019, \mnras, 489,
  4090.
\newblock \doarXiv{1909.01382}

\bibitem[{{Lanzoni} \& {Ciotti}(2003)}]{lanzoni03}
{Lanzoni}, B., \& {Ciotti}, L. 2003, A\&A, 404, 819

\bibitem[{{Lapi} {et~al.}(2018){Lapi}, {Pantoni}, {Zanisi}, {Shi}, {Mancuso},
  {Massardi}, {Shankar}, {Bressan}, \& {Danese}}]{lapi18}
{Lapi}, A., {Pantoni}, L., {Zanisi}, L., {et~al.} 2018, \apj, 857, 22.
\newblock \doarXiv{1803.04734}

\bibitem[{{Magrini} {et~al.}(2012){Magrini}, {Sommariva}, {Cresci}, {Sani},
  {Galametz}, {Mannucci}, {Petropoulou}, \& {Fumana}}]{magrini12}
{Magrini}, L., {Sommariva}, V., {Cresci}, G., {et~al.} 2012, MNRAS, 426, 1195.
\newblock \doarXiv{1206.1513}

\bibitem[{{Maiolino} \& {Mannucci}(2019)}]{maiolino19}
{Maiolino}, R., \& {Mannucci}, F. 2019, \aapr, 27, 3.
\newblock \doarXiv{1811.09642}

\bibitem[{{Maiolino} {et~al.}(2012){Maiolino}, {Gallerani}, {Neri}, {Cicone},
  {Ferrara}, {Genzel}, {Lutz}, {Sturm}, {Tacconi}, {Walter}, {Feruglio},
  {Fiore}, \& {Piconcelli}}]{maiolino12}
{Maiolino}, R., {Gallerani}, S., {Neri}, R., {et~al.} 2012, MNRAS, 425, L66.
\newblock \doarXiv{1204.2904}

\bibitem[{{Maraston} {et~al.}(2010){Maraston}, {Pforr}, {Renzini}, {Daddi},
  {Dickinson}, {Cimatti}, \& {Tonini}}]{maraston10}
{Maraston}, C., {Pforr}, J., {Renzini}, A., {et~al.} 2010, \mnras, 407, 830.
\newblock \doarXiv{1004.4546}

\bibitem[{{Maraston} \& {Str{\"o}mb{\"a}ck}(2011)}]{maraston11}
{Maraston}, C., \& {Str{\"o}mb{\"a}ck}, G. 2011, MNRAS, 418, 2785.
\newblock \doarXiv{1109.0543}

\bibitem[{{Marchesini} {et~al.}(2010){Marchesini}, {Whitaker}, {Brammer}, {van
  Dokkum}, {Labb{\'e}}, {Muzzin}, {Quadri}, {Kriek}, {Lee}, {Rudnick}, {Franx},
  {Illingworth}, \& {Wake}}]{marchesini10}
{Marchesini}, D., {Whitaker}, K.~E., {Brammer}, G., {et~al.} 2010, ApJ, 725,
  1277.
\newblock \doarXiv{1009.0269}

\bibitem[{{Marchesini} {et~al.}(2014){Marchesini}, {Muzzin}, {Stefanon},
  {Franx}, {Brammer}, {Marsan}, {Vulcani}, {Fynbo}, {Milvang-Jensen}, {Dunlop},
  \& {Buitrago}}]{marchesini14}
{Marchesini}, D., {Muzzin}, A., {Stefanon}, M., {et~al.} 2014, \apj, 794, 65.
\newblock \doarXiv{1402.0003}

\bibitem[{{Marsan} {et~al.}(2015){Marsan}, {Marchesini}, {Brammer}, {Stefanon},
  {Muzzin}, {Fern{\'a}ndez-Soto}, {Geier}, {Hainline}, {Intema}, {Karim},
  {Labb{\'e}}, {Toft}, \& {van Dokkum}}]{marsan15}
{Marsan}, Z.~C., {Marchesini}, D., {Brammer}, G.~B., {et~al.} 2015, ApJ, 801,
  133.
\newblock \doarXiv{1406.0002}

\bibitem[{{Mateus} {et~al.}(2007){Mateus}, {Sodr{\'e}}, {Cid Fernandes}, \&
  {Stasi{\'n}ska}}]{mateus07}
{Mateus}, A., {Sodr{\'e}}, L., {Cid Fernandes}, R., \& {Stasi{\'n}ska}, G.
  2007, MNRAS, 374, 1457

\bibitem[{{Mehta} {et~al.}(2018){Mehta}, {Scarlata}, {Capak}, {Davidzon},
  {Faisst}, {Hsieh}, {Ilbert}, {Jarvis}, {Laigle}, {Phillips}, {Silverman},
  {Strauss}, {Tanaka}, {Bowler}, {Coupon}, {Foucaud}, {Hemmati}, {Masters},
  {McCracken}, {Mobasher}, {Ouchi}, {Shibuya}, \& {Wang}}]{mehta18}
{Mehta}, V., {Scarlata}, C., {Capak}, P., {et~al.} 2018, \apjs, 235, 36.
\newblock \doarXiv{1711.05280}

\bibitem[{{Miller} \& {Owen}(2002)}]{miller02}
{Miller}, N.~A., \& {Owen}, F.~N. 2002, \aj, 124, 2453.
\newblock \doarXiv{astro-ph/0207662}

\bibitem[{{Moustakas} {et~al.}(2005){Moustakas}, {Kennicutt}, {Zaritsky}, \&
  {AGN Galaxy Evolution Survey Collaboration}}]{moustakas05}
{Moustakas}, J., {Kennicutt}, R.~C., J., {Zaritsky}, D., \& {AGN Galaxy
  Evolution Survey Collaboration}. 2005, in American Astronomical Society
  Meeting Abstracts, Vol. 207, American Astronomical Society Meeting Abstracts,
  43.02

\bibitem[{{Muzzin} {et~al.}(2013){Muzzin}, {Marchesini}, {Stefanon}, {Franx},
  {Milvang-Jensen}, {Dunlop}, {Fynbo}, {Brammer}, {Labb{\'e}}, \& {van
  Dokkum}}]{muzzin13}
{Muzzin}, A., {Marchesini}, D., {Stefanon}, M., {et~al.} 2013, ApJS, 206, 8.
\newblock \doarXiv{1303.4410}

\bibitem[{{Naab} {et~al.}(2009){Naab}, {Johansson}, \& {Ostriker}}]{naab09}
{Naab}, T., {Johansson}, P.~H., \& {Ostriker}, J.~P. 2009, ApJ, 699, L178.
\newblock \doarXiv{0903.1636}

\bibitem[{{Naab} {et~al.}(2007){Naab}, {Johansson}, {Ostriker}, \&
  {Efstathiou}}]{naab07}
{Naab}, T., {Johansson}, P.~H., {Ostriker}, J.~P., \& {Efstathiou}, G. 2007,
  ApJ, 658, 710

\bibitem[{{Newman} {et~al.}(2018){Newman}, {Belli}, {Ellis}, \&
  {Patel}}]{newman18}
{Newman}, A.~B., {Belli}, S., {Ellis}, R.~S., \& {Patel}, S.~G. 2018, ApJ, 862,
  126.
\newblock \doarXiv{1806.06815}

\bibitem[{{Nipoti} {et~al.}(2009){Nipoti}, {Treu}, {Auger}, \&
  {Bolton}}]{nipoti09}
{Nipoti}, C., {Treu}, T., {Auger}, M.~W., \& {Bolton}, A.~S. 2009, ApJ, 706,
  L86.
\newblock \doarXiv{0910.2731}

\bibitem[{{Nipoti} {et~al.}(2012){Nipoti}, {Treu}, {Leauthaud}, {Bundy},
  {Newman}, \& {Auger}}]{nipoti12}
{Nipoti}, C., {Treu}, T., {Leauthaud}, A., {et~al.} 2012, MNRAS, 422, 1714.
\newblock \doarXiv{1202.0971}

\bibitem[{{Oser} {et~al.}(2012){Oser}, {Naab}, {Ostriker}, \&
  {Johansson}}]{oser12}
{Oser}, L., {Naab}, T., {Ostriker}, J.~P., \& {Johansson}, P.~H. 2012, ApJ,
  744, 63.
\newblock \doarXiv{1106.5490}

\bibitem[{{Oser} {et~al.}(2010){Oser}, {Ostriker}, {Naab}, {Johansson}, \&
  {Burkert}}]{oser10}
{Oser}, L., {Ostriker}, J.~P., {Naab}, T., {Johansson}, P.~H., \& {Burkert}, A.
  2010, ApJ, 725, 2312.
\newblock \doarXiv{1010.1381}

\bibitem[{{Pantoni} {et~al.}(2019){Pantoni}, {Lapi}, {Massardi}, {Goswami}, \&
  {Danese}}]{pantoni19}
{Pantoni}, L., {Lapi}, A., {Massardi}, M., {Goswami}, S., \& {Danese}, L. 2019,
  \apj, 880, 129.
\newblock \doarXiv{1906.07458}

\bibitem[{{Peng} {et~al.}(2015){Peng}, {Maiolino}, \& {Cochrane}}]{peng15}
{Peng}, Y., {Maiolino}, R., \& {Cochrane}, R. 2015, Nature, 521, 192.
\newblock \doarXiv{1505.03143}

\bibitem[{{Pietrinferni} {et~al.}(2004){Pietrinferni}, {Cassisi}, {Salaris}, \&
  {Castelli}}]{pietrinferni04}
{Pietrinferni}, A., {Cassisi}, S., {Salaris}, M., \& {Castelli}, F. 2004, ApJ,
  612, 168

\bibitem[{{Poggianti} \& {Barbaro}(1997)}]{poggianti97}
{Poggianti}, B.~M., \& {Barbaro}, G. 1997, A\&A, 325, 1025.
\newblock \doarXiv{astro-ph/9703067}

\bibitem[{{Renzini}(2006)}]{renzini06}
{Renzini}, A. 2006, ARA\&A, 44, 141

\bibitem[{{Salpeter}(1955)}]{salpeter55}
{Salpeter}, E.~E. 1955, ApJ, 121, 161

\bibitem[{{Saracco} {et~al.}(2020){Saracco}, {Gargiulo}, {La Barbera},
  {Annunziatella}, \& {Marchesini}}]{saracco20}
{Saracco}, P., {Gargiulo}, A., {La Barbera}, F., {Annunziatella}, M., \&
  {Marchesini}, D. 2020, MNRAS, 491, 1777.
\newblock \doarXiv{1911.01438}

\bibitem[{{Saracco} {et~al.}(2019){Saracco}, {La Barbera}, {Gargiulo},
  {Mannucci}, {Marchesini}, {Nonino}, \& {Ciliegi}}]{saracco19}
{Saracco}, P., {La Barbera}, F., {Gargiulo}, A., {et~al.} 2019, MNRAS, 484,
  2281

\bibitem[{{Saracco} {et~al.}(2009){Saracco}, {Longhetti}, \&
  {Andreon}}]{saracco09}
{Saracco}, P., {Longhetti}, M., \& {Andreon}, S. 2009, MNRAS, 392, 718.
\newblock \doarXiv{0810.2795}

\bibitem[{{Saracco} {et~al.}(2011){Saracco}, {Longhetti}, \&
  {Gargiulo}}]{saracco11}
{Saracco}, P., {Longhetti}, M., \& {Gargiulo}, A. 2011, MNRAS, 412, 2707.
\newblock \doarXiv{1011.5740}

\bibitem[{{Saulder} {et~al.}(2015){Saulder}, {van den Bosch}, \&
  {Mieske}}]{saulder15}
{Saulder}, C., {van den Bosch}, R. C.~E., \& {Mieske}, S. 2015, A\&A, 578,
  A134.
\newblock \doarXiv{1503.05117}

\bibitem[{{Schreiber} {et~al.}(2018){Schreiber}, {Glazebrook}, {Nanayakkara},
  {Kacprzak}, {Labb{\'e}}, {Oesch}, {Yuan}, {Tran}, {Papovich}, {Spitler}, \&
  {Straatman}}]{schreiber18}
{Schreiber}, C., {Glazebrook}, K., {Nanayakkara}, T., {et~al.} 2018, \aap, 618,
  A85.
\newblock \doarXiv{1807.02523}

\bibitem[{{Scodeggio} {et~al.}(2005){Scodeggio}, {Franzetti}, {Garilli},
  {Zanichelli}, {Paltani}, {Maccagni}, {Bottini}, {Le Brun}, {Contini},
  {Scaramella}, {Adami}, {Bardelli}, {Zucca}, {Tresse}, {Ilbert}, {Foucaud},
  {Iovino}, {Merighi}, {Zamorani}, {Gavignaud}, {Rizzo}, {McCracken}, {Le
  F{\`e}vre}, {Picat}, {Vettolani}, {Arnaboldi}, {Arnouts}, {Bolzonella},
  {Cappi}, {Charlot}, {Ciliegi}, {Guzzo}, {Marano}, {Marinoni}, {Mathez},
  {Mazure}, {Meneux}, {Pell{\`o}}, {Pollo}, {Pozzetti}, \&
  {Radovich}}]{scodeggio05}
{Scodeggio}, M., {Franzetti}, P., {Garilli}, B., {et~al.} 2005, PASP, 117,
  1284.
\newblock \doarXiv{astro-ph/0409248}

\bibitem[{{Serven} {et~al.}(2005){Serven}, {Worthey}, \& {Briley}}]{serven05}
{Serven}, J., {Worthey}, G., \& {Briley}, M.~M. 2005, \apj, 627, 754

\bibitem[{{Silverman} {et~al.}(2009){Silverman}, {Lamareille}, {Maier},
  {Lilly}, {Mainieri}, {Brusa}, {Cappelluti}, {Hasinger}, {Zamorani},
  {Scodeggio}, {Bolzonella}, {Contini}, {Carollo}, {Jahnke}, {Kneib}, {Le
  F{\`e}vre}, {Merloni}, {Bardelli}, {Bongiorno}, {Brunner}, {Caputi},
  {Civano}, {Comastri}, {Coppa}, {Cucciati}, {de la Torre}, {de Ravel},
  {Elvis}, {Finoguenov}, {Fiore}, {Franzetti}, {Garilli}, {Gilli}, {Iovino},
  {Kampczyk}, {Knobel}, {Kova{\v{c}}}, {Le Borgne}, {Le Brun}, {Mignoli},
  {Pello}, {Peng}, {Perez Montero}, {Ricciardelli}, {Tanaka}, {Tasca},
  {Tresse}, {Vergani}, {Vignali}, {Zucca}, {Bottini}, {Cappi}, {Cassata},
  {Fumana}, {Griffiths}, {Kartaltepe}, {Koekemoer}, {Marinoni}, {McCracken},
  {Memeo}, {Meneux}, {Oesch}, {Porciani}, \& {Salvato}}]{silverman09}
{Silverman}, J.~D., {Lamareille}, F., {Maier}, C., {et~al.} 2009, ApJ, 696,
  396.
\newblock \doarXiv{0810.3653}

\bibitem[{{Springel} {et~al.}(2018){Springel}, {Pakmor}, {Pillepich},
  {Weinberger}, {Nelson}, {Hernquist}, {Vogelsberger}, {Genel}, {Torrey},
  {Marinacci}, \& {Naiman}}]{springel18}
{Springel}, V., {Pakmor}, R., {Pillepich}, A., {et~al.} 2018, MNRAS, 475, 676.
\newblock \doarXiv{1707.03397}

\bibitem[{{Straatman} {et~al.}(2014){Straatman}, {Labb{\'e}}, {Spitler},
  {Allen}, {Altieri}, {Brammer}, {Dickinson}, {van Dokkum}, {Inami},
  {Glazebrook}, {Kacprzak}, {Kawinwanichakij}, {Kelson}, {McCarthy},
  {Mehrtens}, {Monson}, {Murphy}, {Papovich}, {Persson}, {Quadri}, {Rees},
  {Tomczak}, {Tran}, \& {Tilvi}}]{straatman14}
{Straatman}, C. M.~S., {Labb{\'e}}, I., {Spitler}, L.~R., {et~al.} 2014, \apjl,
  783, L14.
\newblock \doarXiv{1312.4952}

\bibitem[{{Tacchella} {et~al.}(2016){Tacchella}, {Dekel}, {Carollo},
  {Ceverino}, {DeGraf}, {Lapiner}, {Mandelker}, \& {Primack}}]{tacchella16}
{Tacchella}, S., {Dekel}, A., {Carollo}, C.~M., {et~al.} 2016, MNRAS, 458, 242.
\newblock \doarXiv{1509.00017}

\bibitem[{{Tanaka} {et~al.}(2017){Tanaka}, {Hasinger}, {Silverman},
  {Bickerton}, {Furusawa}, {Harikane}, {Hu}, {Ikeda}, {Li}, {McCracken},
  {Price}, {Strauss}, {Koike}, {Komiyama}, {Mineo}, {Miyazaki}, {Nishizawa},
  {Takata}, {Utsumi}, {Yamada}, \& {Yasuda}}]{tanaka17}
{Tanaka}, M., {Hasinger}, G., {Silverman}, J.~D., {et~al.} 2017, arXiv
  e-prints, arXiv:1706.00566.
\newblock \doarXiv{1706.00566}

\bibitem[{{Tanaka} {et~al.}(2019){Tanaka}, {Valentino}, {Toft}, {Onodera},
  {Shimakawa}, {Ceverino}, {Faisst}, {Gallazzi}, {G{\'o}mez-Guijarro}, {Kubo},
  {Magdis}, {Steinhardt}, {Stockmann}, {Yabe}, \& {Zabl}}]{tanaka19}
{Tanaka}, M., {Valentino}, F., {Toft}, S., {et~al.} 2019, ApJL, 885, L34.
\newblock \doarXiv{1909.10721}

\bibitem[{{Tanaka} {et~al.}(2020){Tanaka}, {Valentino}, {Toft}, {Onodera},
  {Shimakawa}, {Ceverino}, {Faisst}, {Gallazzi}, {G{\'o}mez-Guijarro}, {Kubo},
  {Magdis}, {Steinhardt}, {Stockmann}, {Yabe}, \& {Zabl}}]{tanaka20}
---. 2020, \apjl, 894, L13

\bibitem[{{Thomas} {et~al.}(2010){Thomas}, {Maraston}, {Schawinski}, {Sarzi},
  \& {Silk}}]{thomas10}
{Thomas}, D., {Maraston}, C., {Schawinski}, K., {Sarzi}, M., \& {Silk}, J.
  2010, MNRAS, 404, 1775.
\newblock \doarXiv{0912.0259}

\bibitem[{{Trager} {et~al.}(2000){Trager}, {Faber}, {Worthey}, \&
  {Gonz{\'a}lez}}]{trager00}
{Trager}, S.~C., {Faber}, S.~M., {Worthey}, G., \& {Gonz{\'a}lez}, J.~J. 2000,
  \aj, 120, 165

\bibitem[{{Trager} {et~al.}(1998){Trager}, {Worthey}, {Faber}, {Burstein}, \&
  {Gonz{\'a}lez}}]{trager98}
{Trager}, S.~C., {Worthey}, G., {Faber}, S.~M., {Burstein}, D., \&
  {Gonz{\'a}lez}, J.~J. 1998, ApJS, 116, 1

\bibitem[{{Valdes} {et~al.}(2004){Valdes}, {Gupta}, {Rose}, {Singh}, \&
  {Bell}}]{valdes04}
{Valdes}, F., {Gupta}, R., {Rose}, J.~A., {Singh}, H.~P., \& {Bell}, D.~J.
  2004, ApJS, 152, 251

\bibitem[{{Valentino} {et~al.}(2020){Valentino}, {Tanaka}, {Davidzon}, {Toft},
  {G{\'o}mez-Guijarro}, {Stockmann}, {Onodera}, {Brammer}, {Ceverino},
  {Faisst}, {Gallazzi}, {Hayward}, {Ilbert}, {Kubo}, {Magdis}, {Selsing},
  {Shimakawa}, {Sparre}, {Steinhardt}, {Yabe}, \& {Zabl}}]{valentino20}
{Valentino}, F., {Tanaka}, M., {Davidzon}, I., {et~al.} 2020, \apj, 889, 93.
\newblock \doarXiv{1909.10540}

\bibitem[{{Valentinuzzi} {et~al.}(2010){Valentinuzzi}, {Fritz}, {Poggianti},
  {Cava}, {Bettoni}, {Fasano}, {D'Onofrio}, {Couch}, {Dressler}, {Moles},
  {Moretti}, {Omizzolo}, {Kj{\ae}rgaard}, {Vanzella}, \&
  {Varela}}]{valentinuzzi10}
{Valentinuzzi}, T., {Fritz}, J., {Poggianti}, B.~M., {et~al.} 2010, ApJ, 712,
  226.
\newblock \doarXiv{0907.2392}

\bibitem[{{van de Sande} {et~al.}(2013){van de Sande}, {Kriek}, {Franx}, {van
  Dokkum}, {Bezanson}, {Bouwens}, {Quadri}, {Rix}, \& {Skelton}}]{vandesande13}
{van de Sande}, J., {Kriek}, M., {Franx}, M., {et~al.} 2013, ApJ, 771, 85.
\newblock \doarXiv{1211.3424}

\bibitem[{{Vazdekis} {et~al.}(1996){Vazdekis}, {Casuso}, {Peletier}, \&
  {Beckman}}]{vazdekis96}
{Vazdekis}, A., {Casuso}, E., {Peletier}, R.~F., \& {Beckman}, J.~E. 1996,
  ApJS, 106, 307.
\newblock \doarXiv{astro-ph/9605112}

\bibitem[{{Vazdekis} {et~al.}(2015){Vazdekis}, {Coelho}, {Cassisi},
  {Ricciardelli}, {Falc{\'o}n-Barroso}, {S{\'a}nchez-Bl{\'a}zquez}, {La
  Barbera}, {Beasley}, \& {Pietrinferni}}]{vazdekis15}
{Vazdekis}, A., {Coelho}, P., {Cassisi}, S., {et~al.} 2015, MNRAS, 449, 1177.
\newblock \doarXiv{1504.08032}

\bibitem[{{Wellons} {et~al.}(2015){Wellons}, {Torrey}, {Ma}, {Rodriguez-Gomez},
  {Vogelsberger}, {Kriek}, {van Dokkum}, {Nelson}, {Genel}, {Pillepich},
  {Springel}, {Sijacki}, {Snyder}, {Nelson}, {Sales}, \&
  {Hernquist}}]{wellons15}
{Wellons}, S., {Torrey}, P., {Ma}, C.-P., {et~al.} 2015, MNRAS, 449, 361.
\newblock \doarXiv{1411.0667}

\bibitem[{{Woo} {et~al.}(2015){Woo}, {Dekel}, {Faber}, \& {Koo}}]{woo15}
{Woo}, J., {Dekel}, A., {Faber}, S.~M., \& {Koo}, D.~C. 2015, MNRAS, 448, 237.
\newblock \doarXiv{1406.5372}

\bibitem[{{Worthey} {et~al.}(1992){Worthey}, {Faber}, \&
  {Gonzalez}}]{worthey92}
{Worthey}, G., {Faber}, S.~M., \& {Gonzalez}, J.~J. 1992, \apj, 398, 69

\bibitem[{{Worthey} \& {Ottaviani}(1997)}]{worthey97}
{Worthey}, G., \& {Ottaviani}, D.~L. 1997, ApJS, 111, 377

\bibitem[{{Yan} {et~al.}(2019){Yan}, {Jerabkova}, \& {Kroupa}}]{yan19}
{Yan}, Z., {Jerabkova}, T., \& {Kroupa}, P. 2019, \aap, 632, A110.
\newblock \doarXiv{1911.02568}

\end{thebibliography}

\end{document}